\documentclass[cmp,10pt]{svjour}
\usepackage{amsmath}
\usepackage{amssymb}
\usepackage{amsfonts}
\usepackage{epsf}
\usepackage{epsfig}
\usepackage{pifont}

\newcommand{\PI}{${\rm P}_{\rm I}\;$}
\newcommand{\PII}{${\rm P}_{\rm II}\;$}
\newcommand{\PIII}{${\rm P}_{\rm III}\;$}

\newcommand{\PV}{${\rm P}_{\rm V}\;$}
\newcommand{\PVI}{${\rm P}_{\rm VI}\;$}
\newcommand{\Z}{\mathbb Z}
\newcommand{\T}{\mathbb T}
\newcommand{\R}{\mathbb R}
\newcommand{\C}{\mathbb C}

\newcommand{\threehalf}{
        {\lower0.00ex\hbox{\raise.6ex\hbox{\the\scriptfont0 3}
                           \kern-.5em\slash\kern-.1em\lower.45ex
                                     \hbox{\the\scriptfont0 2}}}}
\newcommand{\half}{
        {\lower0.00ex\hbox{\raise.6ex\hbox{\the\scriptfont0 1}
                           \kern-.5em\slash\kern-.1em\lower.45ex
                                     \hbox{\the\scriptfont0 2}}}}
\newcommand{\quarter}{
        {\lower0.00ex\hbox{\raise.6ex\hbox{\the\scriptfont0 1}
                           \kern-.5em\slash\kern-.1em\lower.45ex
                                     \hbox{\the\scriptfont0 4}}}}
\newcommand{\eighth}{
        {\lower0.00ex\hbox{\raise.6ex\hbox{\the\scriptfont0 1}
                           \kern-.5em\slash\kern-.1em\lower.45ex
                                     \hbox{\the\scriptfont0 8}}}}
\newcommand{\thirtytwo}{
        {\lower0.00ex\hbox{\raise.6ex\hbox{\the\scriptfont0 1}
                           \kern-.5em\slash\kern-.1em\lower.45ex
                                     \hbox{\the\scriptfont0 32}}}}
\newcommand{\third}{
        {\lower0.00ex\hbox{\raise.6ex\hbox{\the\scriptfont0 1}
                           \kern-.5em\slash\kern-.1em\lower.45ex
                                     \hbox{\the\scriptfont0 3}}}}
\newcommand{\fivehalves}{
        {\lower0.00ex\hbox{\raise.6ex\hbox{\the\scriptfont0 5}
                           \kern-.5em\slash\kern-.1em\lower.45ex
                                     \hbox{\the\scriptfont0 2}}}}
\newcommand{\fourfifths}{
        {\lower0.00ex\hbox{\raise.6ex\hbox{\the\scriptfont0 4}
                           \kern-.5em\slash\kern-.1em\lower.45ex
                                     \hbox{\the\scriptfont0 5}}}}

\renewcommand{\thetable}{\arabic{table}}
\renewcommand{\thefigure}{\arabic{figure}}

\begin{document}

\title{\large\bf 
Painlev\'e transcendent evaluations of finite system
density matrices for 1d 
impenetrable Bosons}

\titlerunning{
Painlev\'e transcendent evaluations of
density matrices}

\date{Received: X X 2002 / Accepted: X X 200X}

\author{
P.J. Forrester\inst{1}, N.E. Frankel\inst{2}, T.M. Garoni\inst{2}
and N.S. Witte\inst{1,2}}

\institute{
Department of Mathematics and Statistics,
University of Melbourne, Victoria 3010, Australia. \\
\email{P.Forrester@ms.unimelb.edu.au, \:
N.Witte@ms.unimelb.edu.au}  \\ 
\and
School of Physics, University of Melbourne, Victoria 3010, Australia. \\
\email{n.frankel@physics.unimelb.edu.au, \:
t.garoni@physics.unimelb.edu.au} }

\maketitle

\communicated{L. Takhtajan}

\begin{abstract}
The recent experimental realisation of a one-dimensional Bose gas of ultra
cold alkali atoms has renewed attention on the theoretical properties of
the impenetrable Bose gas. Of primary concern is the ground state occupation of
effective single particle states in the finite system, and thus the tendency
for Bose-Einstein condensation. This requires the computation of the density 
matrix. For the impenetrable Bose gas on a circle we evaluate the 
density matrix 
in terms of a particular Painlev\'e VI transcendent in $\sigma$-form, and
furthermore show that the density matrix satisfies a recurrence relation in the
number of particles. For the impenetrable Bose gas in a harmonic trap,
and with Dirichlet or Neumann boundary conditions, we give a 
determinant form for the density matrix, a form as an average over the 
eigenvalues of an ensemble of random matrices, and in  special 
cases an evaluation in terms of a transcendent related to Painlev\'e V
and VI. We 
discuss how our results can be used to compute the ground state
occupations. 
\end{abstract}


\section{Introduction}
\setcounter{equation}{0}
Recent advances in the experimental physics of Bose-Einstein condensates
\cite{GVLRGACGIRK_2001,GBMHE_2001,DHRAESPSKSL_2001} have led to the experimental
realisation of a one-dimensional Bose gas of ultra-cold alkali atoms. 
One expects \cite{Ol_1998} that the microscopic
forces are such that there is an effective one-body confining harmonic
potential acting on each atom individually, and an effective infinitely short
range contact potential acting between neighbouring atoms. Moreover, in a
certain physical regime depending on the ratio of the transverse
confinement width to the $s$-wave scattering length, it is argued in
\cite{Ol_1998} that the contact potential can be well approximated by the delta
function form $U(|x-y|) = g\delta(|x-y|)$, and furthermore $g \to \infty$
in the low energy scattering limit. The limit $g \to \infty$ of the delta
function interaction Bose gas is the impenetrable Bose gas, introduced
in \cite{Gi_1960},\cite{LL_1963}.

Not surprisingly, there has thus been renewed interest in the theoretical 
properties of the ground state of the finite system
impenetrable Bose gas \cite{Ol_1998,GWT_2001}. With the 3d Bose gas exhibiting
Bose-Einstein condensation, a central question is the tendency of the finite
system confined to 1d to form a Bose-Einstein condensate. To attack this
question is a two step process. First, with the particles confined to
the region $\Omega \in \R$ and the ground state wave function
$\psi_0$ real, it is necessary to compute the one-particle density
matrix
\begin{equation}
  \rho_N(x;y) = N \int_{\Omega}dx_2 \ldots \int_{\Omega}dx_N
                  \psi_0(x,x_2,\dots,x_N) \psi_0(y,x_2,\dots,x_N),
\label{rho_defn}
\end{equation}
Second, one must solve the eigenvalue problem
\begin{equation}\label{1.2}
   \int_{\Omega} \rho_N(x;y) \phi_k(y) \, dy = \lambda_k  \phi_k(x), 
   \quad k \in \Z_{\geq 0} .
\end{equation}
Because this integral operator is idempotent, the $\lambda_j$ are
non-negative, while the trace condition $\int_{\Omega}
\rho_N(x;x) \, dx = N$ implies $\sum_k \lambda_k = N$. Consequently
the $\lambda_k$ have the interpretation as occupation numbers of
effective single particle states $\phi_k(x)$. The simplest case is when
$\Omega = [0,L]$ with periodic boundary conditions. The periodicity
implies that $\rho_N(x;y) = \rho_N(x-y;0)$. Thus we have $\phi_k(x) =
{1 \over \sqrt{L}} e^{2 \pi i k x/L}$ and so
\begin{equation}\label{1.3}
\lambda_k = \int_0^L \rho_N(x;0) e^{2 \pi i k x /L} \, dx.
\end{equation}
However for other geometries and confinements
there is no analogue of (\ref{1.3}) and one
must solve (\ref{1.2}) numerically.

A number of results are available on $\rho_N(x;0)$ for periodic
boundary conditions. In particular Lenard \cite{Le_1964} has given
$\rho_{N+1}(x;0)$ as an $N \times N$ Toeplitz determinant
(see (\ref{ibg_Cwave})--(\ref{Cmatrix}) below), and subsequently obtained
the $N \to \infty$ asymptotic expansion \cite{Le_1972}
\begin{equation}
   \rho_{N}(x;0) \sim
   \rho_0 A\left({\pi \over N\sin(\pi\rho_0 x/N)} \right)^{1/2},
   \quad A = {G^4(3/2) \over \sqrt{2\pi}}
\label{rho_exp}
\end{equation}
where $ \rho_0 $ denotes the bulk density and $ G(x) $ denotes the Bairn's
G-function, valid for $ x/N $ fixed.
Although the analysis of \cite{Le_1972}
leading to (\ref{rho_exp}) was not rigorous, the setting of the problem as
belonging to the asymptotics of Toeplitz determinants with symbols having
zeros on $ [0,2\pi) $ was identified, and this work inspired a subsequent
rigorous proof \cite{Wi_1973}. (We remark that
the asymptotic form of Toeplitz determinants
of this type was first conjectured by Fisher and Hartwig
\cite{FH_1968}, \cite{FH_1969}. ) The result (\ref{rho_exp})
substituted into (\ref{1.3}) gives 
$\lambda_0 \sim c \sqrt{N}$ for
a specific $ c $ computable from (\ref{rho_exp}).  Thus for large $N$
the fraction of particles in the zero momentum state is proportional to
$\sqrt{N}$.  The result (\ref{rho_exp}) can also be used to compute the
large $ N $ behaviour of $ \lambda_k $ for any fixed $ k \geq 0 $
\cite{FFGW_2002b}.
For the impenetrable Bose gas confined by a harmonic one-body potential,
or indeed in other geometries such as Dirichlet or Neumann boundary
conditions, no results of this type are known.
All one has is the recent
numerical study of Girardeau et al.~\cite{GWT_2001}
in the case of the harmonic well, who by a Monte Carlo study of
system sizes
up to $N=10$ obtained the 
estimate $\lambda_0 \propto
N^{0.59}$ for large $N$. If correct, this result implies the maximum 
effective single particle state  occupation is dependent on the
geometry/confining potential.

To further study this issue, we take up the first step in the procedure
above to compute the $\lambda_j$, and thus provide formulas suitable for the
numerical computation of $\rho_N(x;y)$. Four cases are considered ---
when the domain is a circle (or equivalently periodic boundary conditions); 
a line with
the particles confined by a harmonic one-body potential; and an interval
with Dirichlet or Neumann boundary conditions. The Toeplitz determinant
formulation in the case of periodic boundary conditions is extended to
Hankel determinant forms for $\rho_N(x;y)$ in the other cases
(Section 2.2), and a formulation for efficient Monte Carlo
evaluations by way of expressing the $\rho_N(x;y)$ as averages over
the eigenvalue probability density function (p.d.f.) of certain matrix
ensembles is given (Section 2.3). We then give a systematic Fredholm type
expansion of $\rho_N(x;y)$ about the density $\rho_N(x;x)$ 
(Section 2.4).

Beginning in Section 3 we address the issue of closed form evaluations
of $\rho_N(x;y)$. In the infinite system there are some celebrated
instances of such evaluations. In particular Jimbo et al.~\cite{JMMS_1980}
related the problem of evaluating $\rho_\infty(x;0)$ to integrable
systems theory, and consequently were able to derive the formula
\begin{equation}\label{TLrho}
\rho_\infty(x;0) = \rho_0 \exp \Big (
\int_0^{\pi \rho_0 x} \sigma_V(t) \, {dt \over t} \Big ),
\end{equation}
where $\sigma_V$ satisfies the non-linear equation 
\begin{equation}
   (x\sigma_V'')^2
   + 4(x\sigma_V'-\sigma_V-1)\Big (x\sigma_V'-\sigma_V+(\sigma_V')^2
\Big ) = 0
\label{TLsigma-form}
\end{equation}
subject to the $ x \to 0 $ boundary condition
\begin{equation}
   \sigma_V(x) \mathop{\sim}\limits_{x \to 0}
   -{x^2 \over 3} + {x^3 \over 3\pi} + {\rm O}(x^4) .
\label{TLsigma-exp}
\end{equation}
The differential equation (\ref{TLsigma-form}) is an example of the
so-called Jimbo-Miwa-Okamoto $\sigma$-form of the Painlev\'e V equation,
the latter being essentially the differential equation obeyed by the
Hamiltonian in the Hamiltonian formulation of \PV \cite{Ok_1987b},
\begin{equation}
   (th''_{\rm V})^2 - (h_{\rm V}-th'_{\rm V}+2(h'_{\rm V})^2)^2
  + 4\prod^4_{k=1}(h'_{\rm V}+v_k) = 0
\label{PV_sigma}
\end{equation}
with $ v_1+v_2+v_3+v_4 = 0 $. Setting
\begin{equation}
  \sigma_{V}(x) + \half = h_{\rm V}(t), \quad x=-{it \over 2}
\label{sigma_hV}
\end{equation}
shows that (\ref{TLsigma-form}) reduces to (\ref{PV_sigma}) with
$ (v_1,v_2,v_3,v_4) = (\half,-\half,\half,-\half) $.
Subsequently the characterisation of $\rho_\infty(x;0)$ in terms of the
solution of a differential equation was extended by Its et al.~\cite{IIKS_1990}
(see also \cite{KBI_1993}) to the characterisation of
$\rho_\infty^{ T}(x;0)$ --- the density matrix of the impenetrable Bose
gas at non-zero temperature $T$, as the solution of coupled 
partial differential equations.

In the same study that (\ref{TLsigma-form}) was obtained, 
Jimbo et al.~evaluated the scaled
probability of an eigenvalue free interval for large GUE random matrices
(random Hermitian matrices) in terms of another particular case of the
$\sigma$-form of \PV. In recent years there has been considerable progress
in the evaluation of probabilities and averages in matrix ensembles in
terms of Painlev\'e transcendents (see e.g. \cite{FW_2002b}). Because of
the close relationship between the density matrix for impenetrable bosons
and gap probabilities in matrix ensembles, the random matrix results can
be used to extend the density matrix Painlev\'e transcendent evaluation
(\ref{TLrho}) to the exact Painlev\'e transcendent of 
$ \rho_N(\iota(x);x) $ in the four cases, 
where $ \iota(x) $ denotes the
image of $x$ reflected about the centre of the system.

We adopt two distinct strategies to obtain the exact evaluations.
In Section 3 we present the first approach where we work directly with the
definition of $\rho_N(x;y)$ on a circle as a multidimensional integral.
It turns out that this integral is one of a general class
which have recently \cite{FW_2002b} been
identified as $\tau$-functions for certain \PVI systems.
We show that our \PVI transcendent evaluation for the finite system
scales to the infinite system result (\ref{TLsigma-form}). 
As well as being a special case of the class of integrals related to \PVI
systems in \cite{FW_2002b}, the multidimensional integral formula for
$\rho_N(x;y)$ on a circle is also a special case of a class of integrals
over the unitary group shown to satisfy integrable recurrence relations in
\cite{AvM_2002}. We will show that these recurrences can alternatively
be derived from orthogonal polynomial theory \cite{Ma_2000}.

Underpinning the second of our strategies is the formulation of Lenard
\cite{Le_1964} which allows $\rho_N(x;y)$ to be expressed in terms of the
Fredholm minor of $1 - \xi K_J$, where $K_J$ is the integral operator
on $J = [x,y]$ with kernel $K$ of Christoffel-Darboux type. It is
this formulation which also underlies the calculation of \cite{JMMS_1980}. The
Fredholm minor in turn can be expressed in terms of the product of the
corresponding Fredholm determinant, and the resolvent kernel $R(s,t)$
evaluated at the endpoints $x,y$ of $J$. These latter two quantities have
been extensively studied in the context of gap probabilities in
random matrix ensembles \cite{TW_1993,TW_1994,WFC_2000,WF_2000},
allowing us to
essentially read off from the existing literature an expression for
$\rho_N(\iota(x);x)$ in terms of Painlev\'e transcendents in each case. This
is done in Section 4.

The significance of our results, from the viewpoint of the theory of the
ground state occupation of single particle states for the impenetrable
Bose gas, and from the viewpoint of the Painlev\'e theory, is discussed in
Section 5.

\section{Formulations of $ \rho_N(x;y) $}
\setcounter{equation}{0}
\subsection{The wave functions}
We will first revise the construction of the ground state wave function
for impenetrable bosons on the circle, on the line with a confining
harmonic potential, and on an interval with Dirichlet or Neumann boundary 
conditions. The wave function and density matrix will be given a superscript
"C", "H", "D" and "N" respectively to distinguish the four cases. 

In general the wave function $ \psi(x_1,\ldots,x_N) $ for impenetrable bosons
must vanish at coincident coordinates,
\begin{equation}
   \psi(x_1,\ldots,x_i,\ldots,x_j,\ldots,x_N) = 0 
   \: \: \text{for}\: x_i=x_j, (i \neq j) ,
\label{ibg_defn}
\end{equation}
and satisfy the free particle Schr\"odinger equation otherwise. But for point
particles without spin the condition (\ref{ibg_defn}) is equivalent to the Pauli
exclusion principle. This means that for any fixed ordering of the particles,
\begin{equation}
   x_1 < x_2 < \ldots < x_N
\label{ibg_config}
\end{equation}
say, there is no distinction between impenetrable bosons and free fermions
\cite{Gi_1960}. Consequently the ground state wave function $ \psi_0 $ can,
for the ordering (\ref{ibg_config}), be constructed out of a Slater 
determinant of distinct single particle states. For other orderings $ \psi_0 $
is constructed from the functional form for the sector (\ref{ibg_config}) by
the requirement that it be a symmetric function of the coordinates.

Consider the case that the particles are confined to a circle of
circumference length $L$. This means we require
\begin{equation}\label{8.3}
  \psi(x_1,\dots,x_i+L,\dots,x_N) = \psi(x_1,\dots,x_i,\dots,x_N)
\end{equation}
for each $i=1,\dots,N$. Constructing a Slater determinant obeying (\ref{8.3})
out of distinct single particle states with zero total momentum and
minimum total energy gives
\begin{align}
  \psi^{\rm C}_0(x_1,\dots,x_N) 
  & = (N!)^{-1/2}L^{-N/2} \left\{
               \begin{array}{ll} \displaystyle
       \det[ e^{2\pi ik x_j / L} 
           ]_{j=1,\dots,N \atop k = -(N-1)/2,\dots,(N-1)/2}
     & N \: \text{odd} \\
  \displaystyle
       \det[ e^{2 \pi i(k+1/2) x_j / L} 
           ]_{j=1,\dots,N \atop k = -N/2,\dots,N/2-1}
     & N \: \text{even}
               \end{array} \right. \nonumber \\
  & = (N!)^{-1/2}L^{-N/2}
      \prod_{1 \le j < k \le N} 2i\sin \pi(x_k-x_j)/L
\label{ibg_Cslater}
\end{align}
where the factor of $ (N!)^{-1/2} $ is included so that 
\begin{equation*}
   \int^L_0 dx_1 \cdots \int^L_0 dx_N
   \left| \psi^{\rm C}_0(x_1,\dots,x_N) \right|^2 = 1 .
\end{equation*}
Excluding the (unitary)
factors of $i$, and recalling (\ref{ibg_config}), we note that
this state is non-negative --- a property which distinguishes the ground
state in Bose systems. By the requirement that the wave function for a
Bose system be symmetrical with respect to interchanges
$x_j \leftrightarrow x_{j'} \: (j \ne j')$ we see immediately from 
(\ref{ibg_Cslater}) that for general ordering of particles
\begin{equation}\label{ibg_Cwave}
  \psi^{\rm C}_0(x_1,\dots,x_N) =
  L^{-N/2} (N!)^{-1/2} \prod_{1 \le j < k \le N}
  2 | \sin \pi (x_k - x_j)/L |.
\end{equation}

In the case of impenetrable bosons on a line with a confining harmonic
potential, we take as the Schr\"odinger operator (in reduced units)
\begin{equation}\label{9.4}
  - \sum_{j=1}^N {\partial^2 \over \partial x_j^2} + \sum_{j=1}^N x_j^2.
\end{equation}
The corresponding normalised single particle eigenstates 
$\{ \phi_k(x) \}_{k=0,1,\dots}$ have the explicit form
\begin{equation}
  \phi_k(x) = {2^{-k} \over c^{\rm H}_k} e^{-x^2/2} H_k(x), \quad
  (c^{\rm H}_k)^2 = \pi^{1/2}2^{-k} k!
\label{H_spstates}
\end{equation}
where $H_k(x)$ denotes the Hermite polynomial of degree $k$.
Forming a Slater determinant from the minimal energy states 
($k=0,1,\dots,N-1$), making use of the Vandermonde determinant formula
\begin{equation}\label{10.1}
  \det [p_{j-1}(x_k) ]_{j,k=1,\dots,N}
  = \det [x_k^{j-1} ]_{j,k=1,\dots,N}
  = \prod_{1 \le j < k \le N} (x_k - x_j)
\end{equation}
for any $\{ p_j(x) \}$ with $p_j(x)$ a monic polynomial of degree $j$,
and arguing as in going from (\ref{ibg_Cslater}) to (\ref{ibg_Cwave}) shows
\begin{equation}
  \psi_0^{\rm H}(x_1,\dots,x_N) = {1 \over C^{\rm H}_N}
  \prod_{j=1}^N e^{-x_j^2/2} \prod_{1 \le j < k \le N} | x_k - x_j|,
  \quad (C^{\rm H}_N)^2 = N!\prod^{N-1}_{l=0}(c^{\rm H}_l)^2 .
\label{ibg_Hwave}
\end{equation}

Finally we consider the case of impenetrable bosons on the interval
$[0,L]$ with Dirichlet or Neumann boundary conditions, 
requiring that the wave function or its derivative vanishes at $ x= 0,L $
respectively. The single particle eigenstates $\{ \phi_k(x) \}$
in these situations are, in increasing order of energy,
\begin{equation*}
  \phi^{\rm D}_k(x) = \sqrt{{2 \over L}} \sin{\pi kx \over L}, \;
  (k=1,2,\dots),
  \quad
  \phi^{\rm N}_k(x) = \left \{\begin{array}{ll} 
 \displaystyle \sqrt{{1 \over L}}, & k=0 \\ \displaystyle
\sqrt{{2 \over L}} \cos{\pi kx \over L}, & 
  k=1,\dots \end{array} \right.
\end{equation*}
Recalling the $ C $ and $ D $ type Vandermonde formulas \cite{Pr_1988}
\begin{align*}
   \det[ z^{k}_j - z^{-k}_j ]_{j,k=1,\ldots,n} 
   & =  \prod^{n}_{j=1}(z_j-z^{-1}_j)
        \prod_{1 \leq j < k \leq n}(z_k-z_j)\left( 1- {1 \over z_j z_k} \right)
   \\
   \det[ z^{k-1}_j + z^{-(k-1)}_j ]_{j,k=1,\ldots,n} 
   & = 2\prod_{1 \leq j < k \leq n}(z_k-z_j)\left( 1- {1 \over z_j z_k} \right)
\end{align*}
we see that the corresponding ground state wave functions are
\begin{multline}
  \psi_0^{\rm D}(x_1,\dots,x_N) 
  = {1 \over \sqrt{N!}}
    \Big| \det[ \phi^{\rm D}_k(x_j) ]_{j,k=1,\dots,N}  \Big| \\
  = {1 \over \sqrt{N!}} \left({1 \over \sqrt{2L}} \right)^N
    \prod^{N}_{l=1} 2\sin(\pi x_l/L)
    \prod_{1 \le j < k \le N} 2| \cos \pi x_k/L - \cos \pi x_j/L |,
\label{ibg_Dwave}
\end{multline}
and
\begin{multline}
  \psi_0^{\rm N}(x_1,\dots,x_N) 
  = {1 \over \sqrt{N!}}
    \Big| \det[ \phi^{\rm N}_{k-1}(x_j) ]_{j,k=1,\dots,N}  \Big| \\
  = {1 \over \sqrt{N!}} {1 \over \sqrt{L}}
\left({1 \over \sqrt{2L}} \right)^{N-1}
    \prod_{1 \le j < k \le N} 2| \cos \pi x_k/L - \cos \pi x_j/L | .
\label{ibg_Nwave}
\end{multline}

\subsection{The density matrix as a determinant}
The density matrix $ \rho_{N+1} $ is defined as an $N$-dimensional 
integral by (\ref{rho_defn}). In the cases of the impenetrable Bose gas 
wave functions of the previous section, this integral can be reduced to
a computationally simpler $N$-dimensional determinant. For the circular
case, this form has already been given by Lenard \cite{Le_1964}. Thus
using the general Heine identity
\begin{multline}
   N! \det\left[ \int^{1/2}_{-1/2}dx\; w(z)z^{k-j} \right]_{j,k=1,\ldots,N}
   \\
   = \int^{1/2}_{-1/2}dx_1 \cdots \int^{1/2}_{-1/2}dx_N
     \prod_{l=1}^{N} w(z_l) \prod_{1 \leq j<k \leq N}|z_j-z_k|^2 ,
   \: z_j := e^{2\pi ix_j/L}
\label{heine}
\end{multline}
we see from (\ref{ibg_Cwave}) and (\ref{rho_defn}) that
\begin{align}
   \rho^{\rm C}_{N+1}(x;0)
   & = {1\over L} \det[ a^{\rm C}_{j-k}(x) ]_{j,k=1,\ldots,N}
   \label{ibg_Cdet} \\
   a^{\rm C}_{l}(x) 
   & := \int^{1/2}_{-1/2} dt\; |e^{2\pi ix/L}+e^{2\pi it}|
                            |1+e^{2\pi it}| e^{2\pi ilt} .
   \label{ibg_Cweight}
\end{align}
Furthermore, the elements $ a^{\rm C}_{l} $ have the explicit evaluation
\cite{Le_1964}
\begin{equation}
\begin{split}
   a^{\rm C}_{0} 
   & =
   {4 \over \pi}\left[ \sin{\pi x \over L}
                       +{\pi \over 2}(1-{2x \over L})\cos{\pi x \over L}
                \right]
   \\
   a^{\rm C}_{\pm 1} 
   & =
   {1 \over \pi}e^{\pm i\pi x/L}\left[ \pi(1-{2 x \over L})+\sin{2\pi x \over L}
                                \right]
   \\
   a^{\rm C}_{\pm m}
   & = 
   {4 \over \pi}{(-1)^{m+2} \over m(m^2-1)}e^{\pm im\pi x/L}
         \left[ \cos{\pi x \over L}\sin{m\pi x \over L}
                -m\sin{\pi x \over L}\cos{m\pi x \over L}
         \right], \: |m| > 1
\end{split}
\label{Cmatrix}
\end{equation}
In particular, it follows that
\begin{align}
\rho^{\rm C}_1(x;0) & = {1 \over L} \label{2.20a} \\
  \rho^{\rm C}_2(x;0) 
  & = {4\over \pi L}[\pi(\half-{x \over L})\cos({\pi x \over L})
                     +\sin({\pi x \over L})]
  \label{rho_2} \\
  \rho^{\rm C}_3(x;0) 
  & =
  {8 \over \pi^2 L}\Big\{2-\half\pi^2(\half-{x \over L})^2
     +3\pi(\half-{x \over L})\sin({\pi x \over L})\cos({\pi x \over L})
  \nonumber \\
  & \qquad\qquad
   +[-\fivehalves+2\pi^2(\half-{x \over L})^2]\cos^2({\pi x \over L})
     +\half\cos^4({\pi x \over L}) \Big\} .
  \label{rho_3} 
\end{align}

An essential ingredient underlying the applicability of (\ref{heine}) is
the factorisation
\begin{equation*}
   \psi^{\rm C}_{0}(x,x_1,\ldots,x_N) 
  = {1 \over \sqrt{N+1}}{1 \over \sqrt{L}}
    \prod^{N}_{j=1} \left( 2|\sin({\pi(x-x_j) \over L})| \right)
    \psi^{\rm C}_0(x_1,\ldots,x_N)
\end{equation*}
observed from (\ref{ibg_Cwave}), and their subsequent use of the 
determinant form in (\ref{ibg_Cslater}) to replace $ \psi^{\rm C}_0 $ 
on the right hand side. Now we observe from (\ref{ibg_Hwave}), 
(\ref{ibg_Dwave}), (\ref{ibg_Nwave}) that 
$ \psi^{\rm H}_0, \psi^{\rm D}_0,\psi^{\rm N}_0 $ in the case of $ N+1 $
particles can similarly be factorised. Using the general identity
\begin{multline*}
   N! \det\left[ \int^{\infty}_{-\infty}dt\; g(t)h_{j-1}(t)h_{k-1}(t)
          \right]_{j,k=1,\ldots,N}
   \\
   = \int^{\infty}_{-\infty}dx_1 \cdots \int^{\infty}_{-\infty}dx_N
	\prod_{l=1}^{N} g(x_l) \left( \det[ h_{j-1}(x_k) ]_{j,k=1,\ldots,N}
                               \right)^2 ,
\label{}
\end{multline*}
(c.f. (\ref{heine})) we thus obtain analogous to 
(\ref{ibg_Cdet},\ref{ibg_Cweight}) the determinant formulae
\begin{align}
  \rho^{\rm H}_{N+1}(x;y) 
  & = {1 \over (c^{\rm H}_{N})^2}
      e^{-x^2/2-y^2/2} \det[ a^{\rm H}_{j,k}(x;y) ]_{j,k=1,\ldots,N}
  \label{ibg_Hdet} \\
  \rho^{\rm D}_{N+1}(x;y) 
  & = {2 \over L}
      \sin{\pi x \over L}\sin{\pi y \over L}
      \det[ a^{\rm D}_{j,k}(x;y) ]_{j,k=1,\ldots,N}
  \\
  \rho^{\rm N}_{N+1}(x;y) 
  & = {1 \over 4L} \det[ a^{\rm N}_{j,k}(x;y) ]_{j,k=1,\ldots,N}
      ,\: (N \geq 1)
\end{align}
where
\begin{align}
   a^{\rm H}_{j,k}(x;y)
   & = {2^{-j-k+2} \over c^{\rm H}_{j-1}c^{\rm H}_{k-1}}
       \int^{\infty}_{-\infty}dt\; |x-t||y-t|H_{j-1}(t)H_{k-1}(t)e^{-t^2}
   \label{Hmatrix} \\
   a^{\rm D}_{j,k}(x;y)
   & = 8\int^1_0 dt\; |\cos{\pi x \over L}-\cos{\pi t}|
                      |\cos{\pi y \over L}-\cos{\pi t}|
                      \sin\pi jt \;\sin\pi kt
   \\
   a^{\rm N}_{j,k}(x;y)
   & = 8\int^1_0 dt\; |\cos{\pi x \over L}-\cos{\pi t}|
                      |\cos{\pi y \over L}-\cos{\pi t}|
                      \cos\pi (j\!-\!1)t \cos\pi (k\!-\!1)t .
\end{align}
To simplify further, we note
\begin{equation}
   |x-t||y-t| = \begin{cases}
                  (x-t)(y-t), \quad t \notin [x,y] \\
                 -(x-t)(y-t), \quad t \in [x,y] ,
                \end{cases}
\label{absolute_sign}
\end{equation}
and similarly with 
$|\cos{\pi x \over L}-\cos{\pi t}||\cos{\pi y \over L}-\cos{\pi t}| $.
Use of such an identity allows $ a^{\rm H}_{j,k}(x;y) $ to be evaluated in
terms of incomplete gamma functions, and $ a^{\rm D}_{j,k}(x;y) $, 
$ a^{\rm N}_{j,k}(x;y) $ in a form similar to (\ref{Cmatrix}).

\subsection{$ \rho_{N+1}(x;y) $ and integrals over the classical groups}
In general, for a many body wave function $ \psi_0 $, $ |\psi_0|^2 $ 
has the interpretation as a multivariable p.d.f. As first observed by 
Sutherland in the cases of $ \psi^{\rm C}_0 $ and $ \psi^{\rm H}_0 $, a
feature of $ |\psi_0|^2 $ for each of the wavefunctions (\ref{ibg_Cwave}),
(\ref{ibg_Hwave}), (\ref{ibg_Dwave}) and (\ref{ibg_Nwave}) is that it
coincides precisely with the multivariate p.d.f.~for particular classes
of random matrices. Thus
\begin{align*}
   |\psi^{\rm C}_0|^2 
   & = {\rm Ev}(U(N))|_{\theta=2\pi x/L} \\
   |\psi^{\rm H}_0|^2 
   & = {\rm Ev}({\rm GUE}_N) \\
   |\psi^{\rm D}_0|^2 
   & = {\rm Ev}(Sp(N))|_{\theta=\pi x/L} \\
   |\psi^{\rm N}_0|^2 
   & = {\rm Ev}(O^+(2N))|_{\theta=\pi x/L}
\end{align*}
where ${\rm Ev}(X) $ denotes the eigenvalue p.d.f. of the ensemble of matrices
$ X $, and $ U(N) $ denotes the unitary group with uniform (Haar) measure,
$ {\rm GUE}_N $ the Gaussian unitary ensemble of $ N\times N $ complex Hermitian
matrices, $ Sp(N) $ the group of symplectic unitary $ 2N\times 2N $ matrices
with Haar measure, and $ O^+(2N) $ denotes the group of real orthogonal 
$ 2N\times 2N $ matrices with determinant $ +1 $ and Haar measure.

Moreover, it follows from the definition (\ref{rho_defn}) of the density
matrix, and the explicit forms of the wave functions, that 
$ \rho_{N+1}(x;y) $ in each of the cases can be written as an average over
$ {\rm Ev}(X) $ for appropriate $ X $. Explicitly
\begin{align}
   \rho^{\rm C}_{N+1}(x;0)
   & = {1 \over L}
       \left\langle \prod^{N}_{l=1}
            |2\sin({\pi x \over L}-{\theta_l \over 2})|
            |2\sin({\theta_l \over 2})| \right\rangle_{{\rm Ev}(U(N))}
   \label{rho_Uint} \\
   \rho^{\rm H}_{N+1}(x;y)
   & = {1 \over (c^{\rm H}_{N})^2} e^{-x^2/2-y^2/2}
       \left\langle \prod^{N}_{l=1} |x-x_l||y-x_l|
            \right\rangle_{{\rm Ev}({\rm GUE}_N)}
   \label{rho_GUEint} \\
  \rho^{\rm D}_{N+1}(x;y)
   & = {2 \over L} \sin{\pi x \over L}\sin{\pi y \over L}
   \nonumber\\
   & \phantom{= {2 \over L}} \times
       \left\langle \prod^{N}_{l=1}
            2|\cos{\pi x \over L}-\cos\theta_l|
            2|\cos{\pi y \over L}-\cos\theta_l| \right\rangle_{{\rm Ev}(Sp(N))}
   \label{rho_Spint} \\
  \rho^{\rm N}_{N+1}(x;y)
   & = {1 \over 2L}
       \left\langle \prod^{N}_{l=1}
            2|\cos{\pi x \over L}-\cos\theta_l|
            2|\cos{\pi y \over L}-\cos\theta_l| 
\right\rangle_{{\rm Ev}(O^+(2N))} .
   \label{rho_Oint}
\end{align}
Because it is straightforward to generate typical members from each of these 
matrix ensembles (see e.g. \cite{rmt_Fo}), and so compute eigenvalues from
the corresponding p.d.f. ${\rm Ev}(X) $, these expressions are well suited to
evaluation via the Monte Carlo method.

For future reference we note that the density matrices for the 
corresponding free Fermi systems are given by the same averages, except
that the absolute value signs are to be removed. In particular 
\begin{equation}
   \rho^{\rm C,FF}_{N+1}(x;0)
  = {1 \over L}\left\langle \prod^{N}_{l=1}
            2\sin({\theta_l \over 2}-{\pi x \over L})
            2\sin({\theta_l \over 2}) \right\rangle_{{\rm Ev}(U(N))} .
\label{rhoFF_Uint}
\end{equation}
Furthermore it is elementary to compute density matrices for free Fermi
systems as sums over single particle states (a consequence of all energy
states below the Fermi surface having occupation unity), and this
implies the explicit evaluation
\begin{equation}
   \rho^{\rm C,FF}_{N+1}(x;0) 
   = {1\over L}{\sin(\pi(N+1)x/L) 
                \over \sin(\pi x/L)} .
\label{rhoFF}
\end{equation}

\subsection{Systematic small-$|x-y|$ expansion of $\rho_N(x;y)$}
According to the definition (\ref{rho_defn}), the density matrix at
coincident points $x=y$ is equal to the particle density. But 
the particle
density for the impenetrable Bose gas is the same as for the corresponding
free Fermi system and thus simple to compute. In the infinite system,
the translational invariance of the state gives that the particle density
is a constant. For this case Lenard \cite{Le_1966} has shown how to make
a systematic expansion of the density matrix $\rho_\infty(x;y)$
about the case of
coincident points $\rho_\infty(x;x)$. 
Here we will present this expansion for finite Bose
gas systems with ground state wave functions of the form
\begin{equation}\label{8.1a}
  {1 \over C} \prod_{l=1}^N g(x_l) \prod_{1 \le j < k \le N} |x_k - x_j|. 
\end{equation}
This form includes the case of the harmonic well (\ref{ibg_Hwave}), and
after the change of variables $\cos \pi x_j/L \mapsto x_j$ in
(\ref{ibg_Dwave}) and (\ref{ibg_Nwave}) also includes the case of
Dirichlet and Neumann boundary conditions.

Following Lenard
\cite{Le_1966}, we note that substituting (\ref{8.1a}) in (\ref{rho_defn}) and
using (\ref{absolute_sign}) shows
\begin{align}\label{10.1b}
  \rho_N(x;y) & = {N g(x) g(y) \over C}
  \Big (\int_\Omega - \xi \int_x^y \Big )dx_2
  \, g^2(x_2) \cdots 
\Big ( \int_\Omega dx_N -  \xi \int_x^y \Big )dx_{N}\, g^2(
x_N)
  \nonumber \\
  & \times
  \prod_{l=2}^N ( x- x_l) ( y - x_l)
  \prod_{2 \le j < k \le N} (x_k - x_j)^2\Big|_{\xi =2}.
\end{align}
One now introduces the Fermi type distribution function
\begin{align}\label{10.1c}
  \rho^{\rm FF}_N(x;y;x_2,\dots,x_n)
  & = {N g(x) g(y) \over C}
  \prod_{l=2}^n g^2(x_l)
  \int_\Omega dx_{n+1}
  \, g^2(x_{n+1}) \cdots  \int_\Omega dx_N \, g^2(x_N)
  \nonumber \\
  & \times
  \prod_{l=2}^N ( x- x_l) ( y - x_l)
  \prod_{2 \le j < k \le N} (x_k - x_j)^2.
\end{align}
(when $n=1$ this corresponds to the free fermion one-body density matrix).
Expanding (\ref{10.1b}) in a power series in $\xi$ and using the
definition (\ref{10.1c}) shows
\begin{equation}\label{10.2a}
  \rho_N(x;y) = \sum_{n=0}^\infty {(-\xi)^n \over n!}
  \int_x^y dx_2 \cdots \int_x^y dx_{n+1}
  \rho^{\rm FF}_N(x;y;x_2,\dots,x_{n+1}) \Big |_{\xi = 2}
\end{equation}
(the summation can be extended to infinity since
$\rho^{\rm FF}(x;y;x_2,\dots,x_n) = 0$ for $n > N$).

Next, let $\{ p_j(x) \}_{j=0,1,\dots}$ be monic polynomials of degree
$j$, orthogonal with respect to the weight function $g^2(x)$. Then
writing the integrand in (\ref{10.1c}) as a product of Slater determinants
using (\ref{10.1}) and making use of the orthogonality of the $ p_j(x) $,
a standard calculation shows
\begin{align}\label{4.13'}
  \rho^{\rm FF}_N(x;y;x_2,\dots,x_n)
  & = \det \left [ \begin{array}{cc} K(x,y) &
  [K(x_j,y)]_{j=2,\dots,n} \\
  {}[K(x,x_k)]_{k=2,\dots,n} &
  [K(x_j,x_k)]_{j,k=2,\dots,n} \end{array} \right ] \nonumber \\
  & =: K \Big ( \begin{array}{cccc} x & x_2 & \cdots & x_n \\
  y & x_2 & \cdots & x_n \end{array} \Big )
\end{align}
where, with ${\cal N}_j := \int_{-\infty}^\infty g^2(x)
(p_j(x))^2 \, dx$,
\begin{eqnarray}\label{10.3e}
  && K(x,y) := g(x) g(y) \sum_{j=0}^{N-1}
  {p_j(x) p_j(y) \over {\cal N}_j } \nonumber \\ &&
 =
  {g(x) g(y) \over {\cal N}_{N-1}}
  {p_N(x) p_{N-1}(y) - p_{N-1}(x) p_N(y) \over x - y}
= \rho_N^{\rm FF}(x;y).
\end{eqnarray}
The equality in (\ref{10.3e}) follows from the Christoffel-Darboux
summation formula, and leads to the name Christoffel-Darboux kernel 
(the latter term is due to a relationship with integral equations;
see Section 4.1) for
(\ref{10.3e}). Hence
\begin{align}\label{10.3e'}
  \rho_N(x;y) & = \sum_{n=0}^\infty {(-\xi)^n \over n!}
  \int_x^y dx_2 \cdots \int_x^y dx_{n+1} \,
  K \Big( \begin{array}{cccc}
             x & x_2 & \cdots & x_{n+1} \\
             y & x_2 & \cdots & x_{n+1}
          \end{array} \Big) \Big|_{\xi = 2}
  \nonumber \\
  & := - {1 \over \xi}
  \Delta_{[x,y]}\big(\begin{array}{c}
                    x \\
                    y
                 \end{array} ;\xi\big) \Big |_{\xi = 2}
\end{align}
where
\begin{equation*}
  \Delta_{[x,y]}\big(\begin{array}{c}
                    a \\
                    b
                 \end{array} ;\xi\big) :=
  \sum_{n=0}^\infty {(-\xi)^{n+1} \over n!}
  \int_x^y dx_2 \cdots \int_x^y dx_{n+1} \,
  K \Big ( \begin{array}{cccc}
             a & x_2 & \cdots & x_{n+1} \\
             b & x_2 & \cdots & x_{n+1}
           \end{array} \Big ).
\end{equation*}
As for $|x-y|$ small each term in
(\ref{10.3e'}) is proportional to successively higher powers of
$|x-y|$, this is the sought systematic small $|x-y|$ expansion of
$\rho_N(x;y)$. We will see in Section 4.1 that the expansion
(\ref{10.3e'}) forms the basis for Painlev\'e transcendent evaluations
of $\rho_N(\iota(x),x)$ in the harmonic well, Dirichlet and Neumann
boundary condition cases.

\section{Jimbo-Miwa-Okamoto $\tau$-functions and orthogonal polynomials}
\setcounter{equation}{0}
In this section we will provide the finite $ N $ analogue of the Jimbo, Miwa,
Mori and Sato
\cite{JMMS_1980} Painlev\'e transcendent evaluation (\ref{TLrho}) of 
$ \rho_{\infty}(x;0) $, by similarly evaluating $ \rho^{\rm C}_{N+1}(x;0) $,
and also presenting a recurrence relation in $ N $ for $ \rho^{\rm C}_{N+1}(x;0) $.
Our Painlev\'e transcendent evaluation of $ \rho^{\rm C}_{N+1}(x;0) $ is in terms of
the solution of the Painlev\'e VI equation in $\sigma$-form. Let us then 
discuss some of the theory relating to this equation.

\subsection{Hamiltonian formulation of \PVI and $\tau$-function sequences}
There are six Painlev\'e equations, labelled \PI -- \PVI. They result (see e.g.
\cite{In_1956}) from the project undertaken by Painlev\'e, Gambier and others
to classify solutions to second order differential equations of the form
$ y'' = R(y',y,t) $, where $ R $ is rational in $ y' $, algebraic in $ y $ and
analytic in $ t $ which are free from movable branch points. It was shown that
the only such equations,excluding those which could be reduced to first order 
equations or to linear second order equations, are \PI -- \PVI.
Our interest is in the \PVI equation, which has the form
\begin{multline}
  q'' =
  \half \left( {1 \over q} + {1 \over q - 1} + {1 \over q - t} \right) (q')^2
  - \left( {1 \over t} + {1 \over t-1} + {1 \over q - t} \right) q' \\
  + {q(q-1)(q-t) \over t^2 (t-1)^2}
    \left( \alpha + \beta{t \over q^2} + \gamma{(t-1) \over (q-1)^2}
                  + \delta{t (t-1) \over (q-t)^2} \right) ,
\label{PVI_ode}
\end{multline}
and its solution, the \PVI transcendent $ q(t) $.
We will see that $ \rho^{\rm C}_{N+1}(x;0) $ can be identified with a $\tau$-function
sequence in the \PVI system.

The \PVI system refers to the Hamiltonian system $ \{q,p;H,t\} $
\begin{equation}
   q' = {\partial H\over \partial p}, \quad
   p' = - {\partial H\over \partial q}
\label{Ham_eom}
\end{equation}
where, with $ \alpha_0+\alpha_1+2\alpha_2+\alpha_3+\alpha_4=1 $,
\begin{multline}
  t(t-1) H = q(q-1)(q-t) p^2  \\
     - [ \alpha_4 (q-1) (q-t) + \alpha_3 q(q-t) + (\alpha_0 - 1) q (q-1) ] p \\
     + \alpha_2(\alpha_1 + \alpha_2) (q-t) .
\label{PVI_H}
\end{multline}
It has been known since the work of Malmquist in the early 1920's \cite{Ma_1922}
that the \PVI equation (\ref{PVI_ode}) results from the Hamiltonian system 
(\ref{Ham_eom}), (\ref{PVI_H}) by eliminating $ p $ and choosing the parameters so that 
\begin{equation*}                                   
   \alpha = \half\alpha^2_1,\: \beta = -\half\alpha^2_4,\: 
   \gamma = \half\alpha^2_3,\: \delta = \half(1-\alpha^2_0) .
\end{equation*}
One sees that the Hamiltonian can be written as an explicit rational function
of the \PVI transcendent and its derivative. This follows from the fact that
with $ H $ given by (\ref{PVI_H}), the first of the Hamilton equations is 
linear in $ p $, so $ p $ can be written as a rational function of 
$ q, q' $ and $ t $. 

The $\tau$-function is defined in terms of the Hamiltonian by
\begin{equation}
   H = {d \over d t} \log \tau(t) .
\label{PVI_tau}
\end{equation}
The utility of being able to identify $ \rho^{\rm C}_{N+1}(x;0) $ as a 
$\tau$-function 
for the \PVI system is that $ H $, and thus by integration of (\ref{PVI_tau})
$ \tau(t) $, can be characterised in terms of a differential equation.

\begin{proposition}\cite{JM_1981b,Ok_1987a}
Rewrite the parameters $ \alpha_0,\ldots,\alpha_4 $ of (\ref{PVI_H}) in favour
of the parameters
\begin{equation}
   b_1 = \half(\alpha_3\!+\!\alpha_4), \: 
   b_2 = \half(\alpha_4\!-\!\alpha_3), \:
   b_3 = \half(\alpha_0\!+\!\alpha_1\!-\!1), \: 
   b_4 = \half(\alpha_0\!-\!\alpha_1\!-\!1), 
\label{PVI_b}
\end{equation}
and introduce the auxiliary Hamiltonian $ h $ by
\begin{align}\label{PVI_haux}
  h & = t(t-1) H + e_2'[\mathbf b] t - {1 \over 2} e_2[\mathbf b] \nonumber \\
    & = t(t-1)H + (b_1 b_3 + b_1 b_4 + b_3 b_4) t
                - {1 \over 2} \sum_{1 \le j < k \le 4} b_j b_k ,
\end{align}
where $e_j'[\mathbf b]$ denotes the $j$th degree elementary symmetric
function in $b_1,b_3$ and $b_4$ while $e_j[\mathbf b]$ denotes the
$j$th degree elementary symmetric function in $b_1, \dots, b_4$.
The auxiliary Hamiltonian satisfies the Jimbo-Miwa-Okamoto $\sigma$-form of 
\PVI
\begin{equation}\label{PVI_sigma}
  h'\Big( t(1-t) h'' \Big)^2
  + \Big(h' [2h - (2t - 1) h'] + b_1 b_2 b_3 b_4 \Big)^2 =
    \prod_{k=1}^4( h' + b_k^2).
\end{equation}
\end{proposition}
A self contained derivation of this result can be found in \cite{FW_2002b}.

One of the main practical consequences of the Hamiltonian formulation is
that it allows for a systematic construction of special solutions via
B\"acklund transformations -- birational mappings which leave the Hamilton
equations formally unchanged \cite{Ok_1987a}. The elementary B\"acklund
transformations form an extended affine Weyl group of type $ D_4^{(1)} $.
By composing certain of these elementary operators, shift operators can be
constructed which have the effect of incrementing the 
$ \mathbf{\alpha} $ parameters by $ \pm 1 $ or $ 0 $. For example, one such
operator of this type, denoted $ T_3 $ in \cite{FW_2002b}, has the action
\begin{equation*}
   T_3\boldsymbol{\alpha}
     = (\alpha_0+1,\alpha_1+1, \alpha_2-1, \alpha_3, \alpha_4)
\end{equation*}
or equivalently, after recalling (\ref{PVI_b})
\begin{equation}\label{2.28}
   T_3\mathbf{b} = (b_1, b_2, b_3+1, b_4) .
\end{equation}
Although $ T_3 $ acting on $ p $ and $ q $ is a non-trivial rational
mapping,when acting on $ H $, $ T_3 $ has the formal action of acting
only on the $ \mathbf{\alpha} $'s,
\begin{equation*}
   T_3 H = H \Big |_{\boldsymbol{\alpha} \mapsto T_3\boldsymbol{\alpha}} .
\end{equation*}
This motivates introducing a sequence of Hamiltonians
\begin{equation*}
   T_3^n H = H \Big |_{\boldsymbol{\alpha} \mapsto T_3^n\boldsymbol{\alpha}} ,
\end{equation*}
and a corresponding sequence of $\tau$-functions specified by
\begin{equation}\label{2.35a}
   T_3^n H = {d \over dt} \log \tau_3[n], \qquad
   \tau_3[n] = \tau_3[n](t) = \tau(t;b_1,b_2,b_3+n,b_4) .
\end{equation}
A crucial result due to Okamoto \cite{Ok_1987a}, which can be derived from 
the specific form of the action of $ T_3 $ and $ T_3^{-1} $ on $ H $, $ p $ and
$ q $ \cite{KMNOY_2001}, is that $ \tau_3[n] $ satisfies a particular 
differential recurrence relation.

\begin{proposition}
The $\tau$-function sequence (\ref{2.35a}) satisfies the Toda lattice equation
\begin{equation}\label{fi}
  \delta^2 \log \bar{\tau}_3[n]
   = { \bar{\tau}_3[n-1]  \bar{\tau}_3[n+1] \over  \bar{\tau}_3^2[n] },
     \qquad \delta = t(t-1) {d \over dt}
\end{equation}
where
\begin{equation}\label{fip}
   \bar{\tau}_3[n]
    := \Big( t(t-1) \Big)^{(n+b_1+b_3)(n+b_3+b_4)/2} {\tau}_3[n].
\end{equation}
\end{proposition}

The significance of this is that an identity of Sylvester (see \cite{Mu_1960})
gives that if
\begin{equation}\label{fi3}
   \bar{\tau}_3[0] = 1,
\end{equation}
then the general solution of (\ref{fi}) is given by
\begin{equation}\label{fi5}
   \bar{\tau}_3[n] 
    = \det \Big[ \delta^{j+k} \bar{\tau}_3[1] \Big]_{j,k=0,1,\dots,n-1}.
\end{equation}
Furthermore, restricting the parameter space so that $ \alpha_2 = 0 $ 
(which corresponds to a chamber wall or reflection hyperplane in the
affine $ D^{(1)}_4 $ root system), it has been shown by Okamoto 
\cite{Ok_1987a} that $ \bar{\tau}_3[1] $ is given in terms of a solution
of the Gauss hypergeometric equation. Using integral solutions of the 
latter, the formula (\ref{fi5}) was taken as the starting point by 
Forrester and Witte \cite{FW_2002b} in an extensive study of multidimensional
integral forms of the $\tau$-function sequence $ \bar{\tau}_3[n] $.
In particular, results relating to averages of the form (\ref{rho_Uint}),
equivalent to the following were established.

\begin{proposition}
Define
\begin{multline}
  A_N(u;\omega,\mu;\xi) = \\
  \Big\langle \prod^{N}_{l=1}(1-\xi\chi^{(l)}_{[0,\phi)})
                              \left( 2\sin{\theta_l \over 2} \right)^{2\omega}
                              \left( {-1 \over ue^{i\theta_l}} \right)^{\mu}
                              (1-ue^{i\theta_l})^{2\mu}
  \Big\rangle_{{\rm Ev}(U(N))}\Big|_{u=e^{-i\phi}}
\label{A_defn}
\end{multline}
where $ z_l = e^{i\theta_l} $, $ 0 \leq \theta_l \leq 2\pi $, and
\begin{equation*}
   \chi^{(l)}_{J} = \begin{cases}
                       1, & \theta_l \in J \\
                       0, & \theta_l \notin J .
                    \end{cases}
\end{equation*}
Let 
\begin{equation}
  \mathbf{b} = \Big( \half(N\!+\!\omega\!-\!\mu),
                     \omega + \half(N\!+\!\omega\!+\!\mu),
                     \half(N\!-\!\omega\!+\!\mu),
                    -\mu-\half(N\!+\!\omega\!+\!\mu) \Big),
\label{A_b_param}
\end{equation}
and write
\begin{equation*}
  C_1 = e_2'[\mathbf{b}] + \mu N, \qquad
  C_2 = \half e_2[\mathbf{b}] + \mu N
\end{equation*}
(recall the definition of $ e_2'[\mathbf{b}] $ and $ e_2[\mathbf{b}] $ from
Proposition 1). The \PVI system with parameters (\ref{A_b_param}) permits
the $ \tau$-function sequence
\begin{equation}
   \tau_3[N] \propto u^{N\mu/2}A_N(u;\omega,\mu;\xi)
\label{tau_Aseq}
\end{equation}
where the proportionality factor is independent of $ u $ and furthermore 
\begin{equation}
   C_{1}u-C_{2} + u(u-1){d \over du}\log A_N(u;\omega,\mu;\xi)
  = h_{\rm VI}(u;\mathbf{b})
\label{A_haux}
\end{equation}
where $ h_{\rm VI}(t;\mathbf{b}) $ is an auxiliary Hamiltonian (\ref{PVI_haux})
for the \PVI system with parameters (\ref{A_b_param}). Consequently 
(\ref{A_haux}) satisfies the \PVI equation in $\sigma$-form (\ref{PVI_sigma})
with parameters (\ref{A_b_param}).
\end{proposition}

To relate (\ref{A_defn}) to (\ref{rho_Uint}) we note that 
\begin{equation*}
  (1-2\chi^{(l)}_{[0,\phi)}) \left( {-1 \over ue^{i\theta_l}} \right)^{1/2}
  (1-ue^{i\theta_l})|_{u=e^{-i\phi}}
  = 2|\sin{(\theta_l-\phi) \over 2}| .
\end{equation*}
Consequently
\begin{equation}
   \rho^{\rm C}_{N+1}(x;0) = {1 \over L}A_N(e^{2\pi ix/L};\half,\half;2)
\label{rho_A}
\end{equation}
where we have used the fact that $ \rho^{\rm C}_{N+1}(x;0) $ is even in $ x $.

The choice of the parameters in (\ref{rho_A}) corresponding to 
$ \rho^{\rm C}_{N+1} $ implies a special structure to the $ \tau$-function 
sequence (\ref{tau_Aseq}). First substituting (\ref{rho_A})
in (\ref{A_b_param}) shows we are considering the \PVI
system with parameters
\begin{equation}
   \mathbf{b} = \big( \half N,1+\half N,\half N,-1-\half N \big) .
\label{rho_b_param}
\end{equation}
As noted above, $ \tau_3[1] $ satisfies the Gauss hypergeometric differential
equation. The parameters in the latter are related to the parameters $ \mathbf{b} $
by
\begin{equation*}
   a=b_1+b_4, \; b=1+b_3+b_4, \; c=1+b_2+b_4 .
\end{equation*}
Substituting the special values (\ref{rho_b_param}) we see that in particular
$ c=1 $, which is the condition for the existence of a logarithmic solution at 
the origin ($ u=0 $). For general $ N $, $ \tau_3[N] $ then corresponds to a
generalisation of this logarithmic solution of the Gauss hypergeometric
equation. To illustrate this point, we note that with $ \mathbf{b} $ given by
(\ref{rho_b_param}), according to (\ref{tau_Aseq}) and (\ref{rho_A}) we have
\begin{equation*}
   \tau_3[N](u) \propto u^{N/2}\rho^{\rm C}_{N+1}(x;0)\Big|_{u=e^{2\pi ix/L}} .
\end{equation*}
Recalling (\ref{rho_2}) and (\ref{rho_3}) we see 
\begin{align*}
   \tau_3[1](u)
  & \propto (u+1)v + 2(u-1)
  \\
   \tau_3[2](u)
  & \propto 4(u^2+u+1)v^2 + 12(u-1)(u+1)v
  \nonumber \\
  & \phantom{\propto}\qquad 
            - u^{-1}(u-1)^2(u^2-14u+1)
\end{align*}
where
$ v = \pi i-\log u $ which exhibits the further structure of being a polynomial
of degree $ N $ in $ v $, and a Laurent polynomial in $ u $ of positive degree
$ N $ and negative degree $ N-1 $. 

The \PVI system with parameters (\ref{rho_b_param}) also permits a $ \tau$-function
sequence which is strictly a polynomial. To anticipate this we relate 
(\ref{A_defn}) to the free Fermi average (\ref{rhoFF_Uint}) by noting
\begin{equation*}                                              
  \left( {-1 \over ue^{i\theta_l}} \right)^{1/2}   
  (1-ue^{i\theta_l})|_{u=e^{-i\phi}}                          
  = 2\sin{(\theta_l-\phi) \over 2} ,
\end{equation*}
and so deducing
\begin{equation}                                                                
   \rho^{\rm C,FF}_{N+1}(x;0) = {1 \over L}A_N(e^{2\pi ix/L};\half,\half;0) .             
\label{rhoFF_A}                                                                  
\end{equation}
Recalling (\ref{tau_Aseq}) and (\ref{rhoFF}) 
we see that this corresponds to the 
$ \tau$-function sequence
\begin{equation*}                                                 
   \tau_3[N](u) \propto \sum^{N-1}_{j=0} u^j .
\end{equation*}
This class of polynomial solutions is a special case of the generalised Jacobi 
polynomial solutions identified in \cite{NOOU_1998}. As a final remark on the 
theme of special classes of solutions to the \PVI system, we note that the
specification of the parameters (\ref{rho_b_param}) is a particular example 
which permits elliptic solutions \cite{KK_1998,DIKZ_1999,Hi_1995}. More generally 
the latter occur when
\begin{equation*}
\begin{split}
   t_1 & = 1+b_3-b_4 = 2+N \in \Z
   \\
   t_2 & = b_1+b_2 = 1+N \in \Z
   \\
   t_3 & = b_1-b_2 = -1 \in \Z
   \\
   t_4 & = 1+b_3+b_4 = 0 \in \Z
\end{split}
   \qquad \sum^{4}_{k=1}t_k = 2(N+1) \in 2\Z .
\end{equation*}

Substituting (\ref{rho_A}) into (\ref{A_haux}) of Proposition 3 
and replacing $N$ by $N-1$ throughout gives the
sought evaluation of $ \rho^{\rm C}_{N}(x;0) $ in terms of a solution of the
\PVI equation in $\sigma$-form.

\begin{corollary}
Define
\begin{equation}
   \sigma_{N}(u) 
   := u(u-1){d \over du}\log\rho^{\rm C}_{N}(x;0)|_{e^{2\pi ix/L}=u}
\label{IBG_sigma_defn}
\end{equation}
so that
\begin{equation}
   \rho^{\rm C}_{N}(x;0) = \rho_0
   \exp\left( 2\pi i\int^{x/L}_0{dt \over e^{2\pi it}-1}\sigma_{N}(e^{2\pi it})
       \right) .
\label{IBG_sigma_defn2}
\end{equation}
The quantity $ \sigma_{N}(u) $ satisfies the particular \PVI $\sigma$-form 
differential equation
\begin{multline}
  u^2(u-1)^2(\sigma_{N}'')^2 \\
     + [\sigma_{N}-(u-1)\sigma_{N}'+1]
     \Big\{ 4\sigma_{N}'(\sigma_{N}-u\sigma_{N}')
            -(N^2-1)[\sigma_{N}-(u-1)\sigma_{N}']
     \Big\} = 0
\label{IBG_sigma_ode}
\end{multline}
subject to the boundary condition
\begin{align}
   \sigma_{N}(u)  \mathop{\sim}\limits_{u \to 1}
   & {N^2-1 \over 12}(u-1)^2
      +{(N^2-1)(iN-\pi) \over 24\pi}(u-1)^3 +\ldots
\label{IBG_sigma_bc}
\end{align}
The formula (\ref{IBG_sigma_ode}) also holds for $ \rho^{\rm C,FF}_{N}(x;0) $,
except that $ \sigma^{\rm FF}_{N}(u) $ is now subject to the boundary condition
\begin{align*}
   \sigma^{\rm FF}_{N}(u)  \mathop{\sim}\limits_{u \to 1} 
   & {N^2-1 \over 12}(u-1)^2
	-{N^2-1 \over 24}(u-1)^3+\ldots
\end{align*}
\end{corollary}

Proof - This is immediate from Proposition 3 and (\ref{rho_A}), (\ref{rhoFF_A}),
except for the boundary conditions. The latter in the free Fermi case follows by
substituting the exact evaluation
(\ref{rhoFF}) in (\ref{IBG_sigma_defn}). Use is also made of the free Fermi
density matrix exact evaluation (\ref{rhoFF}) to deduce the boundary
condition in the impenetrable Bose gas case. Thus according to
(\ref{4.13'})--(\ref{10.3e'}) we have
\begin{eqnarray}\label{IBG_rho_bc}
  \rho^{\rm C}_{N}(x;0) & = & \rho^{\rm C \, FF}_{N}(x;0) - 2
  \int_0^x \det \left | \begin{array}{cc}
  \rho^{\rm C\, FF}_{N}(x;0) & \rho^{\rm C\,FF }_{N}(x_2;0) \\
  \rho^{\rm C\, FF}_{N}(x;x_2) & \rho^{\rm C\, FF}_{N}(x_2;x_2)
  \end{array} \right | \, dx_2 + \cdots \nonumber \\
  & \mathop{\sim}\limits_{x \to 0} &
  \rho_0 \Big ( 1 - {(N-1)(N+1) \over 6} \Big ( {\pi x \over L}
  \Big )^2 + {(N-1)N(N+1) \over 9\pi}\left({\pi x\over L}\right)^3
     + \ldots
  \Big ), \nonumber \\
\end{eqnarray}
where the second line follows after substituting (\ref{rhoFF}) and expanding
the first term to $O(x^2)$ (this term only contains even powers of $x$),
and the second term to its leading order, $O(x^3)$. Finally we substitute
(\ref{IBG_rho_bc}) in (\ref{IBG_sigma_defn}) to deduce the expansion
(\ref{IBG_sigma_bc}).
\hfill{$\square$}

One immediate consequence of Corollary 1 is that it allows the small $ x $ 
expansion to easily be extended. Thus it follows that the corrections to
(\ref{IBG_rho_bc}) at order $ x^4 $ and $ x^5 $ are 
\begin{equation}
     +{(N-1)(N+1)[3N^2-7] \over 360}\left({\pi x\over L}\right)^4
     -{(N-1)N(N+1)[11N^2-29] \over 1350\pi}\left({\pi x\over L}\right)^5
\label{IBG_rho2_bc}
\end{equation}

The results (\ref{TLrho}), (\ref{TLsigma-form}) of Jimbo et al \cite{JMMS_1980} 
follow simply from our results (\ref{IBG_sigma_defn2}), (\ref{IBG_sigma_ode}). 
Thus defining 
$ \sigma_V(t) = \lim_{N \to \infty} \sigma_{N}(e^{2it/N}) $
we obtain (\ref{TLrho}) from (\ref{IBG_sigma_defn2}), while substituting 
$ u = e^{2it/N} $ in (\ref{IBG_sigma_ode}), replacing
$ \sigma_{N}(e^{2it/N}) $ with $ \sigma_V(t) $ and equating the leading 
order terms in $N$ (which are $O(1)$)
to zero gives (\ref{TLsigma-form}). The boundary condition 
(\ref{TLsigma-exp}) corresponds to the scaled limit of (\ref{IBG_sigma_bc}).

\subsection{Orthogonal polynomials on the unit circle}
A feature of a number of recent studies 
\cite{FW_2001a,FW_2002b,AvM_2002,Bo_2001,BB_2002} relating Hankel and Toeplitz 
determinants to
Painlev\'e transcendents has been the characterisation of the former
not only as the solution of nonlinear differential equations, but also
as the solution of nonlinear difference equations. Here we will show
a difference equation characterisation is also possible for
$\rho_N^{\rm C}(x;0)$.

For this purpose we adopt an orthogonal polynomial approach, similar to
that used in \cite{IW_2001}.
The characterisation of the density matrix as a Toeplitz determinant with
a non-negative and bounded symbol (\ref{ibg_Cdet},\ref{ibg_Cweight})
immediately implies an underlying orthogonal polynomial system 
defined on the unit circle. The weight appearing in (\ref{ibg_Cweight}) is
the special case $a=b=1/2$ of the generalised Jacobi weight 
\begin{equation}
   w(z) = {C \over 2\pi}|1+z|^{2a}|1+uz|^{2b}, \quad a,b \in \C, \;
   z \in \mathbb{T}
\label{gJ_weight}
\end{equation}
where $C$ is the normalisation
\begin{equation}\label{normC}
   {C \over 2\pi}\int_{\T}{dz \over iz}|1+z|^{2a}|1+uz|^{2b} = 1 .
\end{equation}
Associated with (\ref{gJ_weight}) is a system of orthonormal
polynomials $ \{ \phi_n(z) \}^{\infty}_{n=0,1,\dots}$,
\begin{equation*}
   \int_{\T}{dz \over iz}w(z)\phi_n(z)\overline{\phi_m(z)} = \delta_{m,n}.
\end{equation*}
In obtaining a recurrence relation for
\begin{equation*}
  D_{N-1} := \det\Big[ \int_{\T}{dz \over iz} w(z)z^{k-j} \Big]_{j,k=1,\ldots,N}, 
\end{equation*} 
and thus since
\begin{equation}\label{sn0}
\rho^{\rm C}_{N+1}(x;0) = {1 \over L C^{N}} D_{N-1} |_{a=b=1/2}
\end{equation}
for the density matrix,
one focuses attention on the leading two coefficients $\kappa_n$,
$l_n$ in $\phi_n(z)$, and the trailing coefficient $\phi_n(0)$,
\begin{equation}\label{3.48}
\phi_n(z) = \kappa_n z^{n} + l_n z^{n-1} + \ldots + \phi_n(0).
\end{equation}
The relevance of $\kappa_n, \phi_n(0)$ are seen from the Szeg\"o relations
\cite{ops_Sz}
\begin{equation*}
   \kappa_n^2 = {D_{n-1} \over D_{n}}, \qquad
\kappa_n^2 = \sum_{k=0}^n | \phi_k(0) |^2
\end{equation*}
which show in particular that 
\begin{equation}\label{sn1}
  1 - |r_N|^2 = { D_{N-2} D_{N} \over D_{N-1}^2}, \qquad
  r_n := {\phi_n(0) \over \kappa_n}.
\end{equation}
We will see that for the weight
(\ref{gJ_weight}),
the Freud equations
-- which are recurrence relations among the successive coefficients 
$ \kappa_n, \phi_n(0) $ --  have a special structure which leads to a
recurrence equation for $r_n$, and thus according to
(\ref{sn0}) and (\ref{sn1}), for $\rho^{\rm C}_N(x;0)$.
\begin{proposition}
Consider the special case $a=b=1/2$ of (\ref{gJ_weight}), in which
case according to (\ref{sn0}) the relation (\ref{sn1}) reads
\begin{equation}\label{3.51}
  1-|r_{N}|^2 = {\rho^{\rm C}_{N+2}\rho^{\rm C}_{N}
                  \over (\rho^{\rm C}_{N+1})^2}.
\end{equation}
The ratios $r_n$, and thus via (\ref{3.51}) the successive density matrices,
are determined by the third order difference equation with respect to
$N$ 
\begin{equation}\label{refl_diff}
\begin{split}
   2\cos{\pi x\over L} + 2\tilde{r}_{N+1}\tilde{r}_{N}
  = & {1-\tilde{r}_{N+1}^2 \over \tilde{r}_{N+1}}
      \left[(N+3)\tilde{r}_{N+2}+(N+1)\tilde{r}_{N}\right]
    \\
    & - {1-\tilde{r}_{N}^2 \over \tilde{r}_{N}}
        \left[(N+2)\tilde{r}_{N+1}+N\tilde{r}_{N-1}\right]
\end{split}
\end{equation}
where $ r_n := e^{i\pi(1-x/L)n}\tilde{r}_n \in \R$.
The initial members of this sequence of $ \tilde{r}_n $ required to start the 
recurrence are
\begin{align}
   \tilde{r}_0 & = 1 \label{refl_0}
   \\
   \tilde{r}_1 
   & = \quarter{\pi - 2\pi x/L + \sin(2\pi x/L) \over 
               {1\over 2}(\pi - 2\pi x/L)\cos(\pi x/L) + \sin(\pi x/L) }
   \label{refl_1} 
\end{align}
(substituting these values in (\ref{refl_diff}) with $N=0$ allows
$\tilde{r}_2$ to be computed). Also, the initial members of the
$\rho_N^{\rm C}$ sequence are specified by (\ref{2.20a}) and
(\ref{rho_2}) (with these values given $\rho_3^{\rm C}$ is computed
from (\ref{3.51}) with $N=1$). 
\end{proposition}

Proof - Magnus has found a recurrence relation \cite{Ma_2000} for the 
ratios $r_n$
applicable to the generalised Jacobi weight (\ref{gJ_weight})
\begin{equation}
   (n+1+a+b)r_{n+1} + (n-1+a+b)\bar{u}r_{n-1}
    = {\bar{u}\bar{l}_n/\kappa_n+l_n/\kappa_n-n(\bar{u}+1) \over 1-|r_n|^2}
\label{M_recur}
\end{equation}
However this involves both $ r_n $ 
and $ l_n $ and we require a further relation to determine $ l_n $. This relation 
is 
\begin{equation}
   {l_n \over \kappa_n}-\bar{u}{\bar{l}_n \over \kappa_n}
   = {(a-b)n \over n+a+b}(\bar{u}-1) ,
\label{W_recur}
\end{equation}
and follows from telescoping the identity
\begin{equation*}
   (n+a+b+1) \left[ {l_{n+1} \over \kappa_{n+1}}
                     -\bar{u}{\bar{l}_{n+1} \over \kappa_{n+1}} \right]
 - (n+a+b) \left[ {l_n \over \kappa_n}-\bar{u}{\bar{l}_n \over \kappa_n} \right]
 = (a-b)(\bar{u}-1) ,
\end{equation*}
which in turn is derived by evaluating
\begin{equation*}
  \int_{\T} {dz \over iz} (1+z)(1+uz)w'(z)\phi_n(z)\overline{\phi_n(z)}
\end{equation*}
in two different ways.
Then (\ref{refl_diff}) follows after setting $ a=b=\half $ and extracting a 
phase factor of $ e^{i\pi(1-x/L)n} $.

The initial conditions (\ref{refl_0}), (\ref{refl_1}) can be determined by a
Gram-Schmidt type construction of the orthonormal polynomials
$\{\phi_n(z)\}_{n=0,1,\dots}$. First, due to the normalisation of the
weight (\ref{normC}) we have $\phi_0(z) = 1$ and thus (\ref{refl_0})
follows. With
\begin{equation*}
   \langle f, g \rangle := \int_{\T}{dz \over iz} \, w(z) \Big|_{a=b=1/2}
   f(z) \overline{g(z)},
\end{equation*}
the orthogonality $\langle \phi_1(z), \phi_0(z) \rangle = 0$, explicit
value  $\phi_0(z) = 1$ and  (\ref{3.48}) give
$\kappa_1 \langle z, 1 \rangle + \phi_1(0) = 0$. The value of
$\langle z, 1 \rangle$ can be read off from (\ref{Cmatrix}), thus implying
(\ref{refl_1}). 
\hfill{$\square$}

We note that the special cases corresponding to $x$ at either the
end points ($x=0,L$ and thus $u=1$) or the midpoint ($x=L/2$, and thus
$u=-1$) allow simple explicit formulas for the $r_n$. Thus we have  
 \cite{IW_2001}
\begin{equation}\label{3.61}
\begin{split}
  u = 1, \quad  & r_N = (-1)^N{1 \over N+1} \\
  u = -1, \quad & r_{N=2p} = {1 \over N+1}, \quad r_{N=2p+1} = 0 \\
\end{split}
\end{equation}
which clearly satisfy (\ref{refl_diff}). The density matrix at these
points also has a closed form evaluation,
\begin{align*}
  \rho_N^{\rm C}(0;0) 
  & = {N \over L}
  \\
  \rho_{N=2p}^{\rm C}(L/2;0)
  & = {1 \over L} \left({4 \over \pi}\right)^{N-1}
      4^{(p-1)(2p-1)} {G(p+2)G^6(p+1)G(p) \over G^2(2p+1)}
  \\
  \rho_{N=2p+1}^{\rm C}(L/2;0)
  & = {1 \over L} \left({4 \over \pi}\right)^{N-1}
      4^{p(2p-1)} {G^4(p+1)G^4(p+2) \over G^2(2p+2)},
\end{align*}
where $ G(x) $, the Barnes G-function, has the explicit form 
$ G(x) = (x-2)!(x-3)! \ldots 1! $ for $ x \in \Z_{\geq 2} $.
Here the former evaluation follows from the general fact that at coincident
points the density matrix is equal to the particle density, while the
latter makes use of results from \cite{IW_2001} on the explicit form of
the $\kappa_n$'s for the weight (\ref{gJ_weight}), 
$ \kappa_{2n} = \kappa_{2n+1} = (2n+1)!/2^{2n}(n!)^2\sqrt{n+1} $,  
in the case $u=-1$, $a=b$. 
The small $ x $ expansion of $ \rho^{\rm C}_{N}(x;0) $, (\ref{IBG_rho_bc})
and (\ref{IBG_rho2_bc}), substituted into (\ref{3.51}) allow the corresponding small
$ x $ expansion of $ \tilde{r}_n $ to be computed up to a sign, which 
in turn can be determined using (\ref{3.61}). This shows 
\begin{equation*}
  \tilde{r}_n \sim (-1)^n \left\{ {1 \over n+1}
    + {n(n+2) \over 6(n+1)}\left({\pi x \over L}\right)^2
    - {n(n+2) \over 3\pi}\left({\pi x \over L}\right)^3 
    + \ldots \right\} .
\end{equation*}

A feature of the explicit forms (\ref{3.61}) is that $|r_N| \to 0$ as
$N \to \infty$. According to (\ref{sn1}) this is a necessary condition
for the convergence of $\rho_N^{\rm C}(x;0)$ as $N \to \infty$. Here we note
that with $|r_N|$ small the difference equation (\ref{refl_diff}) simplifies
to read
\begin{equation}\label{Bn}
2 \cos {\pi x \over L} = B_{N+1} - B_N, \qquad
B_N := {(N+2) \tilde{r}_{N+1} + N \tilde{r}_{N-1} \over \tilde{r}_N}.
\end{equation}
It follows from (\ref{Bn})
that $B_N = 2 N \cos \pi x / L + C$, where $C$ is independent of
$N$. Noting that this implies
$B_N \sim 2 (N+1) \cos \pi x / L$ for $N$ large, we thus obtain the
recurrence
\begin{equation}
  (N+2) \tilde{r}_{N+1} + N \tilde{r}_{N-1} =
  (N+1) \tilde{r}_{N} 2\cos \pi x / L,
\end{equation}
which with $A_N := (N+1) \tilde{r}_N$, $ \cos \pi x /L := t$ reads
\begin{equation}\label{An}
  A_{N+1} + A_{N-1} = 2t A_N.
\end{equation}
This is precisely the three term recurrence satisfied by the
Chebyshev polynomials \cite{ops_Sz}). Thus (\ref{refl_diff}) can be
regarded as a non-linear generalisation of the Chebyshev recurrence
(\ref{An}).

Although it is not obvious from the derivation, the equations
(\ref{3.51}) and (\ref{refl_diff}) remain valid in the free Fermi
case. This can be seen by substituting the exact evaluation
(\ref{rhoFF}) into (\ref{3.51}) to deduce
\begin{equation*}
  \tilde{r}_n = {\sin \pi x / L \over \sin \pi (n+1) x / L}
\label{refl_FF}
\end{equation*}
and then verifying that this is an exact solution of (\ref{refl_diff}).
Unlike $ \tilde{r}_n $ in the Bose case, (\ref{refl_FF}) does not
obey the inequality $ |\tilde{r}_n| < 1 $ for all $ x $.
 
As our final point of the difference equation, we remark that
recently Adler and van Moerbeke \cite{AvM_2002} have constructed essentially 
the same pair of coupled recurrences (\ref{M_recur}), (\ref{W_recur}) 
from their 
theory of the Toeplitz lattice and its Virasoro algebra. In the particular case at 
hand their weight is 
specialised to $ \alpha = \beta = 1 $, and $ \xi^{-2} = u = e^{2\pi ix/L} $.
Their variables are related to ours by $ x_n = \tilde{r}_n $ and through the
use of their relation (0.0.14), which is the analogue of (\ref{W_recur}), then 
one can show $ y_n = x_n $. The other recurrence in their work, (0.0.15) is the
analogue of (\ref{M_recur}) and can be shown to lead to  
\begin{multline*}
   (1-x^2_n)\left[ (n+2)x_{n+1}x_{n-1} + n+1 \right]
  -(1-x^2_{n-1})\left[ (n-1)x_{n}x_{n-2} + n \right] \\
  = 1 + 2x_{n-1}x_n\cos(\pi x/L) + x^2_{n-1}x^2_n \\
    -1 +(1-x^2_1)(3x_2+2)-x_1(x_1+2\cos(\pi x/L)) .
\end{multline*}
Now by clearing denominators and rearranging (\ref{refl_diff}) one can recover
the first five terms of the above relation. Furthermore by using the initial
conditions of the 
recurrence (\ref{refl_0}), (\ref{refl_1}) one can show that the
sum of the last three terms is identically zero and thus the two forms are the
same.

\section{Painlev\'e-type evaluations of $ \rho_N(\iota(x);x) $}
\setcounter{equation}{0}
\subsection{Fredholm formulation}
While the unitary average (\ref{rho_Uint}) defining $ \rho^{\rm C}_{N+1}(x;0) $ 
is a known
$\tau$-function in the Painlev\'e theory, the same is not true of the 
averages (\ref{rho_GUEint}) -- (\ref{rho_Oint}). Indeed the density matrices
in these cases are genuinely functions of both $x$ and $y$. These variables
play the role of time in the Hamiltonian formulations of the Painlev\'e 
equations, so there being more than one time variable, we are taken outside
this class. However, with $ y = \iota(x) $, where $ \iota(x) $ denotes the
reflection of $ x $ about the centre of the system (thus $ \iota(x) = -x $
for the harmonic well case, and $ \iota(x) = L-x $ for the case of the
Dirichlet and Neumann boundary conditions) we again have a function of one
variable.
Although this cannot be recognised as a single $\tau$-function, it turns
out that we can formulate the calculation of $\rho_N(\iota(x);x)$ so that
it is expressed in terms of quantities known in terms of Painlev\'e
transcendents from random matrix theory. For this one makes use of
a classical operator theoretic interpretation of (\ref{10.3e'}) relating
to Fredholm integral equations \cite{Le_1966}. 

It is the latter formulation which has
been used in the pioneering work of Jimbo et al.~\cite{JMMS_1980}
on the evaluation of the bulk density matrix in terms of a
Painlev\'e V transcendent, and the generalisation of this result by
Its, Korepin and coworkers \cite{IIKS_1990,KBI_1993} to the temperature
dependent bulk density matrix. The key point is that with $K_J$
denoting the integral
operator on $J=[x,y]$ with kernel (\ref{10.3e}), and 
$R(a,b;\xi)$ denoting the kernel of the resolvent operator
$R := \xi K_J(1 - \xi K_J)^{-1}$, it is true in general that 
(see e.g.~\cite{Le_1966,JMMS_1980})
\begin{equation}\label{4.17}
  \Delta_{[x,y]}\big(\begin{array}{c}
                    a \\
                    b
                 \end{array} ;\xi\big) = - \xi \det(1 - \xi K_J) R(a,b;\xi)
\end{equation}
(the quantity $\Delta_{[x,y]}$ is called the first Fredholm minor).
Now in the harmonic well case and the cases of Dirichlet and Neumann
boundary conditions (the latter two after the change of variables
$\cos \pi x_j/L \mapsto x_j$) the wave function is of the form
(\ref{8.1a}) with
\begin{equation}\label{4.18}
g^2(x) = \left \{ \begin{array}{ll} e^{-x^2}, & {\rm harmonic \; well} \\
(1-x^2)^{\pm 1/2}, &
{\rm Dirichlet \: and \: Neumann} \end{array} \right.
\end{equation}
These weights have the property of being even is $x$. This implies a
special structure to (\ref{4.17}) if $J$ is also chosen to be symmetrical
about the origin, $J = [-x,x]$ say. Thus a consequence of $g^2(x)$ being
even is that the orthogonal polynomials $p_j(x)$ are even for
$j$ even and odd for $j$ odd, and this from
(\ref{10.3e}) implies $K(a,b) = K(-a,-b)$. Using this latter property,
and with $J = [-x,x]$, it is true in general that (see e.g.~\cite{TW_1993})
\begin{equation*}
  {d \over dx} \log \det (1 - \xi K_J) = - 2 R(x,x;\xi).
\end{equation*}
Antidifferentiating and substituting in (\ref{4.17}) with
$[x,y] \mapsto [-x,x]$, then substituting the result in (\ref{10.3e'})
shows
\begin{equation}
  \rho_N(-x;x) =
   R(-x,x;\xi) \exp\Big( -2\int_0^x R(t,t;\xi) \,dt \Big) \Big|_{\xi=2}.
\label{R0_inner}
\end{equation}

The crucial point of the formula (\ref{R0_inner}) is that the quantities
$R(-x,x)$ and $R(t,t)$, for Christoffel-Darboux kernels corresponding to
the weights (\ref{4.18}) have previously been
calculated in terms of Painlev\'e transcendents as part of studies into gap
probabilities (interval $J$ free of eigenvalues)
for random matrix ensembles, the GUE in the case of the
harmonic well, and the JUE with $a=b=\pm 1/2$ in the case of Dirichlet
and Neumann boundary conditions. Although (\ref{4.18}) and (\ref{R0_inner})
have general validity, the specific integrable nature of the kernel (\ref{10.3e})
\cite{IIKS_1990} is essential for this characterisation.  
In the latter case the quantities in
(\ref{R0_inner}) were studied in \cite{WFC_2000}, but with
$J= (-1,-x] \cup [x,1)$ rather than $J=[-x,x]$. To overcome
this difference in detail, we note we can rewrite (\ref{10.1b}) to read
\begin{multline*}
  \rho_N(x;y) = (1-\xi)^{N-1} {N g(x) g(y) \over C}
  \left( \int_\Omega + {\xi \over 1 - \xi} \Big(
  \int_{-\infty}^{x} + \int_y^{\infty} \Big) \right) dx_2 \, g^2(x_2) \\
  \cdots
  \left( \int_\Omega + {\xi \over 1 - \xi} \Big(
  \int_{-\infty}^{x} + \int_y^{\infty} \Big) \right) dx_N \, g^2(x_N) \\
  \times
  \prod_{l=2}^N(x-x_l)(y-x_l) 
  \prod_{2 \le j < k \le N} (x_k - x_j)^2 \Big |_{\xi = 2}.
\end{multline*}
Repeating the working which led to (\ref{R0_inner}) then shows
\begin{equation}
  \rho_N(-x;x) = (-1)^{N-1} R(-x,x;\xi)
  \exp \Big (-2 \int_x^\infty R(t,t;\xi) \, dt \Big ) \Big |_{\xi = 2}
\label{R0_outer}
\end{equation}
where $R$ now denotes the kernel of the resolvent operator
$R = \xi K_{\bar{J}}(1 - \xi K_{\bar{J}})^{-1}$,
$\bar{J} := (-\infty,-x] \cup [x,\infty)$.

\subsection{Evaluation of $\rho_N^{\rm H}(-x;x)$ and 
$\rho_N^{\rm D, N}(L-x;x)$}
Let us begin by specifying the quantities in (\ref{R0_inner}) in
the harmonic well case. From the above discussion, this corresponds to the 
interval $J$ being eigenvalue free in the GUE. For this matrix
ensemble, the gap probability for both the interval $J=[-x,x]$, and
the interval $J=(-\infty,-x] \cup [x,\infty)$ have been studied
\cite{TW_1994,WFC_2000}, allowing us to use either (\ref{R0_inner}) 
or (\ref{R0_outer}) to deduce an exact expression for 
$\rho^{\rm H}_{N}(-x;x)$. For purposes of specifying the boundary
condition in the corresponding differential equation, it is most
convenient to use the latter.

\begin{proposition}
For the impenetrable Bose gas on a line, confined by a harmonic well,
we have
\begin{equation*}
  \rho^{\rm H}_{N}(-x;x)
    = \tilde{R}(x;\xi) \exp \Big( - 2 \int_x^\infty R(t;\xi) \, dt
                            \Big) \Big|_{\xi = 2} 
\end{equation*}
where with $h:= \sqrt{s^2 - 2 R'}$, $R$ satisfies the equation
\begin{equation*}
  s R'' + 2R' = 2s (s-h) - 2h
  \sqrt{(R+sR')^2 - 4s^2(s-h)R - 2Ns^2(s-h)^2}
\end{equation*}
while $\tilde{R}$ satisfies the equation
\begin{multline*}
  \Big( s\tilde{R}'' + 2\tilde{R}' + 8Ns\tilde{R} - 24s^2\tilde{R}^2
  \Big)^2 \\
  = 4(s - 2\tilde{R})^2 \Big( (\tilde{R}+s\tilde{R}')^2
        + 8Ns^2\tilde{R}^2 - 16s^3\tilde{R}^3 \Big).
\end{multline*}
The corresponding boundary conditions  are
\begin{align*}
  R(s;\xi)   \mathop{\sim}\limits_{s \to \infty} \tilde{R}(s;\xi)
  & \mathop{\sim}\limits_{s \to \infty}
    (\xi-1)^{N-1} \rho^{\rm H}_{N}(s;s)
  \\
  & \mathop{\sim}\limits_{s \to \infty}
  (\xi-1)^{N-1}{2^{N-1} \over \pi^{1/2} (N-1)!} s^{2N-2}e^{-s^2}
\end{align*}
\end{proposition}
\hfill{$\square$}

The two resolvent kernels are not independent being related by 
\begin{equation*}
   {d \over ds}R(s;\xi) = -2s\tilde{R}(s;\xi)-2\tilde{R}^2(s;\xi)
\end{equation*}
so that $ h = s + 2\tilde{R}(s;\xi) $.
Both resolvent kernels have been reduced to a particular \PV transcendent $ w(x) $
with parameters
\begin{equation*}
   \alpha = \eighth N^2, \quad \beta = -\eighth(N-\epsilon)^2, \quad
   \gamma = \half\epsilon, \quad \delta = -\half,
\end{equation*}
where $ \epsilon = \pm 1 $. The reductions were found to be
\begin{align*}
   R & = {1 \over 8\sqrt{x}w(w-1)^2}
     \left\{ 2x{d \over dx}w+N(w-1)^2+(2x-1)w+1 \right\}
     \nonumber \\
     & \phantom{= {1 \over 8\sqrt{x}w(w-1)^2}} \times
     \left\{ 2x{d \over dx}w-N(w-1)^2-(2x+1)w+1 \right\}
     \\
   \tilde{R} & = 
     -{\sqrt{x} \over 2w(w-1)} \left[ \epsilon{d \over dx}w - w \right]
     + {N(w-1)+\epsilon \over 4\sqrt{x}w}
\end{align*}
where $ s^2 = x $.

In the case of Dirichlet and Neumann boundary conditions we require
$R(-t,t)$ and $R(t,t)$ for the symmetric JUE with
$ (-1,-t] \cup [t,1) $ eigenvalue free. The differential equation
satisfied by $R(t,t)$ is known from \cite{WFC_2000}, but the equation
for $R(-t,t)$ was not made explicit in that work. We therefore give
some details of the required calculation below.

\begin{proposition}
The impenetrable Bose gas on the finite interval $ [0,L] $ subject to 
Dirichlet or Neumann boundary conditions at the ends has the density matrix
\begin{equation*}
   \rho^{\rm D,N}_{N}(L-x;x) = {\pi \over L}
\sin {\pi x \over L} R^{\rm D,N}_0(s;\xi) 
   \exp\left( -2\int^{1}_{s} dt R^{\rm D,N}(t;\xi) \right)\Big|_{\xi=2
\atop s = \cos \pi x /L}
\end{equation*}
where $ \sigma(s) := (1-s^2)R^{\rm D,N}(s;\xi) $ satisfies 
\begin{multline}\label{sigmaODE_DN}
  \left\{ 
   {s(1-s^2) \over F}\left[ (N+\alpha)^2s + 2s\sigma' - (1-s^2)\sigma'' 
                     \right] + (1+s^2)F + 2(N+\alpha)s
    \right\}^2 \\
  - \left\{ 
    2(1-s^2)\sigma - 2(N+\alpha)s^2[(N+\alpha)s+F] -s[(N+\alpha)s+F]^2
    \right\}^2 \\
  = -4s^2[(N+\alpha)s+F]^2 \left\{ 
      N(N+2\alpha) + 2s\sigma + 2(N+\alpha)s[(N+\alpha)s+F] \right\}
\end{multline}
where $ \alpha = \pm \half $ and $ F := \sqrt{(N+\alpha)^2s^2-2(1-s^2)\sigma'} $
and with the boundary condition
\begin{align}\label{sigmabc_DN}
  R^{\rm D,N}(s;\xi) 
  & \mathop{\sim}\limits_{s \to 1} 
     (\xi-1)^{N-1}\rho^{\rm D,N}_{N}(s;s)
  \\
  & \mathop{\sim}\limits_{s \to 1} 
    (\xi-1)^{N-1}{ (N+\alpha)\Gamma(N+2\alpha+1) \over 
      2^{\alpha+1}(N-1)!\Gamma(\alpha+1)\Gamma(\alpha+2) }(1-s)^{\alpha}
\end{align}
and $ R_0 := R^{\rm D,N}_0(s;\xi) $ satisfies 
\begin{multline}\label{R0ODE_DN}
   \Big\{ s(1-s^2)R''_0 + 2(1-s^2)(1-2s^2)R'_0 \\
          + 8sR_0(2sR_0-(N+\alpha)/2)(2sR_0-N-\alpha) - 2(1-s^2+2\alpha^2)sR_0
   \Big\}^2 \\
   = 4\left\{-(N+\alpha)s + 2(1+s^2)R_0 \right\} \\
      \times\Big\{ (1-s^2)^2(R_0+sR'_0)^2 
                   + 4(sR_0)^2[(2sR_0-N-\alpha)^2 - \alpha^2] \Big\}
\end{multline}
with the boundary condition
\begin{align}\label{R0bc_DN}
  R^{\rm D,N}_0(s;\xi) 
  & \mathop{\sim}\limits_{s \to 1} 
    (\xi-1)^{N-1} K^{\rm D,N}_{N}(-s,s)
  \\
  & \mathop{\sim}\limits_{s \to 1} 
    (\xi-1)^{N-1}
    { \Gamma(N+2\alpha+1) \over 
      2^{\alpha+1}(N-1)!\Gamma^2(\alpha+1) }(1-s)^{\alpha}
\end{align}
\end{proposition}

Proof - Both cases come under the symmetric Jacobi weight discussed in
\cite{WFC_2000} however this study doesn't contain sufficient details for our
purposes. From Proposition 8 of the above we can derive an integral of the
system of differential equations. Combining (4.26) and (4.27) we have 
\begin{equation}\label{DN_9}
  (1-s^2)(qp)' - [\beta_0+u(2\alpha_1\!-\!1)]p^2 
               + [\gamma_0-w(2\alpha_1\!+\!1)]q^2 = 0
\end{equation}
and we can rewrite (4.23) as
\begin{equation}\label{DN_10}
   \sigma - [\beta_0+u(2\alpha_1\!-\!1)]p^2 - [\gamma_0-w(2\alpha_1\!+\!1)]q^2
   -2\alpha_1 sqp - 2s^{-1}(1-s^2)q^2p^2 = 0 .
\end{equation}
Adding (\ref{DN_10}) to $ 2\alpha_1 $ times (\ref{DN_9}) and employing 
(4.24,25,28) we find the sum to be a perfect derivative, and given that 
$ R_0, \sigma, u, w $ all vanish as $ s \to 1 $  the integral is
\begin{equation*}
   [\beta_0+u(2\alpha_1\!-\!1)][\gamma_0-w(2\alpha_1\!+\!1)]
   -2s\sigma - 4\alpha_1(1-s^2)qp = \beta_0\gamma_0 = N(N+2\alpha) .
\end{equation*}
Given this integral we can follow the procedure in the proof of Proposition 11
in \cite{WF_2000} to derive second-order differential equations for $ \sigma $ 
and $ R_0 $, and equations (\ref{sigmaODE_DN}), (\ref{R0ODE_DN}) are the results.
The boundary conditions are found from the kernels 
\begin{align*}
   K_N(t,t)
   & = {N!\Gamma(N+2\alpha+1) \over 2^{2\alpha+1}\Gamma^2(N+\alpha)}
   (1-t^2)^{\alpha}
   \nonumber \\
   & \qquad\times \left[
    P^{(\alpha+1,\alpha)}_{N-1}(t)P^{(\alpha,\alpha+1)}_{N-1}(t)
    - P^{(\alpha,\alpha)}_{N}(t)P^{(\alpha+1,\alpha+1)}_{N-2}(t) \right]
   \\
   K_N(-t,t) 
   & = (-)^{N-1}{N!\Gamma(N+2\alpha+1) \over 
                 2^{2\alpha+1}\Gamma(N+\alpha)\Gamma(N+\alpha+1)}
   {(1-t^2)^{\alpha} \over t}
     P^{(\alpha,\alpha)}_{N-1}(t)P^{(\alpha,\alpha)}_{N}(t)
\end{align*}
where the $ P^{(\alpha,\beta)}_{N}(t) $ is the standard Jacobi polynomial.
\hfill{$\square$}

Again the two resolvent kernels are connected by
\begin{equation*}
   {d \over ds}\sigma = -2(N+\alpha)sR_0 - 2(1-s^2)R^2_0
\end{equation*}
This system was solved in terms of \PVI transcendent
$w(x)$ with parameters
\begin{equation*}
   \alpha = \eighth, \quad \beta = -\half(N+\alpha-\epsilon/2)^2, \quad
   \gamma = 0, \quad \delta = \half(1-\alpha^2) 
\end{equation*}
according to 
\begin{align*}
   R_0(s) 
  & = {\epsilon \over 2\sqrt{x}w}{d \over dx}w
      - {1 \over 4\sqrt{x}(x-1)}{w-1 \over w}
        \left[ \epsilon(w+x)-2(N+\alpha)x \right]
  \\
  \sigma(s)
  & = -{\left[ 2(x-1){\displaystyle d \over \displaystyle dx}w
                - (w-1)(w+x) \right]^2 \over 8\sqrt{x}w(w-1)(w-x) }
  \nonumber \\
  & \qquad 
      - {\sqrt{x}(w-1) \over 2w(w-x)}
        \left[ N(N+2\alpha)w-(N+\alpha)^2x \right] 
\end{align*}
where $ s = \sqrt{x} $ and $ \epsilon = \pm 1 $. 

We remark that the thermodynamic form of $\rho_N^{\rm D, N}(x;y)$ for 
$x$ and $y$ fixed but general
has been studied in the spirit of the paper of
Jimbo et al.~by Kojima \cite{Ko_1997}, 
although the final characterisations obtained
(in terms of an integrable differential system) does not appear to be
amenable to computation.

\subsection{Thermodynamic limit}
In the thermodynamic limit, Jimbo et al \cite{JMMS_1980} were able to
identify the first Fredholm minor directly as a $\tau$-function, and so had
no need for the formula (\ref{R0_inner}). Nonetheless it can be used to
evaluate $ \rho_{\infty}(x;0) $ in terms of a solution of (\ref{PV_sigma}),
as we will now demonstrate. First, repeating the working which lead to
(\ref{R0_inner}) shows that with $ K_J $ the integral operator on $ J=(-x,x) $
with kernel $ K(x,y) = \sin(x-y)/\pi(x-y) $ and
$ R = \xi K_J(1-\xi K_J)^{-1} $,
\begin{equation*}
   \rho_{\infty}(-x;x) = \pi\rho_0 R_{\infty}(-\pi\rho_0 x, \pi\rho_0 x;\xi)
   \exp\left( -2\int^{\pi\rho_0 x}_{0} R_{\infty}(t,t;\xi) dt
       \right)\Big|_{\xi=2} .
\end{equation*}
Furthermore, we know from \cite{JMMS_1980} that 
\begin{equation}
   -{t\over 2}R_{\infty}(\quarter\; t,\quarter\; t;\xi)
  = h_{\rm V}(-it;(0,0,0,0)) ,
\label{tr5}
\end{equation}
where $ h_{\rm V}(t;\bf{v}) $ satisfies (\ref{PV_sigma}), subject to the
boundary condition
\begin{equation*}
   h_{\rm V}(-it;(0,0,0,0)) \mathop{\sim}\limits_{t \to 0}
   -\xi{t \over 2\pi} - \xi^2{t^2 \over 4\pi^2} ,
\end{equation*}
and we know too that
\begin{equation*}                                                   
   {d \over dt}R_{\infty}(t,t;\xi) = 2(R_{\infty}(-t,t;\xi))^2
\end{equation*}
(see e.g. \cite{TW_1993}). Consequently $ \rho_{\infty}(x;0) $ can be 
expressed in terms of the transcendent (\ref{tr5}) according to
\begin{multline}
  \rho_{\infty}(x;0) = \pi\rho_0
  \left( -{d\over dt}{h_{\rm V}(-it;(0,0,0,0)) \over t} \right)^{1/2} \\
  \times \exp\left( \int^t_0 {h_{\rm V}(-iu;(0,0,0,0)) \over u} du 
             \right)\Big|_{t=2\pi\rho_0 x \atop \xi=2} .
\label{sr1}
\end{multline}
This is to be compared against the evaluation (\ref{TLrho}) due to Jimbo et
al \cite{JMMS_1980}, with the substitution (\ref{sigma_hV}) and the boundary
condition (\ref{TLsigma-exp}) generalised to be consistent with (\ref{10.3e'}),
\begin{gather}
   \rho_{\infty}(x;0) = \rho_0
   \exp \int^t_0{-\half+h_{\rm V}(-iu;(\half,-\half,\half,-\half)) \over u}du
        \Big|_{t=2\pi\rho_0 x \atop \xi=2} ,
   \\
   h_{\rm V}(-it;(\half,-\half,\half,-\half)) \mathop{\sim}\limits_{t \to 0}
   {1\over 2} + {t^2 \over 12} + {i\xi t^3 \over 48\pi} . 
\label{sr2}
\end{gather}
Equating the logarithmic derivatives of (\ref{sr1}) and (\ref{sr2}) gives the
identity
\begin{multline*}
  h_{\rm V}(s;(\half,-\half,\half,-\half)) \\
  = \half + h_{\rm V}(s;(0,0,0,0)) 
          + {s\over 2}{d\over ds}\log\left( -{d\over ds}
            {h_{\rm V}(s;(0,0,0,0)) \over s} \right) .
\end{multline*}
It remains to derive this directly from the Painlev\'e theory. In
this regard we note that an identity with similar characteristics
for the \PII system can be deduced from a classical result of
Gambier, while the analogous result for the \PIII system has only
recently been found \cite{Wi_2002}.

\section{Discussion}
\setcounter{equation}{0}
We will conclude with a discussion of our results. In relation to 
$ \rho^{\rm C}_{N}(x;0) $ the recurrence relation (\ref{refl_diff}) allows 
rapid and stable tabulation for very large values of $ N $ for all 
$ x \in [0,L) $. Hence numerical evaluations of the $\lambda_k$ in
(\ref{1.3}) can be carried out. 
Although for fixed $k$ the leading behaviour in $N$ of $\lambda_k$ are
known from (\ref{rho_exp}), of interest is the convergence of the 
$ \lambda_k $ (appropriately scaled) to their thermodynamic value. 
Furthermore, the differential equation characterisation of 
$ \rho^{\rm C}_{N}(x;0) $ given in Corollary 1 is well suited to generating 
the power series expansion about $ x=0 $. This relates to the behaviour of 
$ \lambda_k $ for $ k $ large. A comprehensive study of such issues will be 
discussed in a forthcoming publication \cite{FFGW_2002b}.

Our results show a remarkable Fermi-Bose correspondence at a mathematical
level of characterising the density. It goes back to Girardeau that up to
the sign under permutation of the coordinates, the ground state wave function 
of the 1d free Fermi and impenetrable Bose systems are identical. This means 
that quantities depending only on the absolute value squared of $ \psi_0 $
are the same for both systems. The density matrix is not of this type, and
so distinguishes the two systems. Nonetheless, we find that the same 
differential and difference equations characterise $ \rho^{\rm C}_{N}(x;0) $
for both the Bose and Fermi systems - they are only distinguished at this
mathematical level by the boundary conditions. For $ \rho_{\infty}(x;0) $
this same property was already observed by Jimbo et al \cite{JMMS_1980}.
One would like to be able to use the differential equation to see how the
different prescribed small $ x $ behaviours imply different behaviours at
large $ x $. In the case of $ \rho_{\infty}(x;0) $, Jimbo et al 
\cite{JMMS_1980} were able to do this and so obtain the expansion
\begin{equation}
   \rho_{\infty}(x;0) \sim
   \rho_0 {A \over \sqrt{\rho_0 x}}
   \left( 1 + {1 \over 8(\pi \rho_0 x)^2}
              \left[ \cos(2\pi \rho_0 x)-\quarter \right] 
            + {\rm O}({1 \over x^4})
   \right),
\label{TLrho_exp}
\end{equation}
which was first derived by Vaidya and Tracy \cite{VT_1979,VT_1979b}. A 
challenging problem is to use Corollary 1 to deduce higher order terms in the 
expansion (\ref{rho_exp}) of $ \rho^{\rm C}_{N}(x;0) $ for large $ N $, 
with $ x/N $ fixed. Note that in the Fermi case this expansion can be read off 
from (\ref{rhoFF}).

For the density matrix $ \rho^{\rm H}_N(x;y) $ 
(and similarly $\rho_{N}^{\rm D,N}(x;y)$)
we have the presented determinant
formulation specified by (\ref{ibg_Hdet}), (\ref{Hmatrix}), as well as the 
formulation
(\ref{rho_GUEint}) as an average over the eigenvalues of the GUE. Both these
forms are suitable for the numerical computation of $ \rho^{\rm H}_N(x;y) $ for 
general $ x $ and $ y $. In the special case that $ (x;y) \mapsto (-x;x) $ we
have given an explicit functional form in terms of the transcendents 
related to \PV. This provides a much more efficient numerical scheme for the
computation of $ \rho^{\rm H}_N(-x;x) $, and will provide a valuable test for
the accuracy of Monte Carlo evaluation via (\ref{rho_GUEint}).

With this achieved, the next step in the determination of the occupation
numbers is  the numerical solution of the integral equation
(\ref{1.2}).
In addition to the numerical evaluation
of the occupations, one would like to determine the exact leading asymptotic
behaviour of $ \lambda_i $, $ i $ fixed, from an asymptotic expression for 
$ \rho^{\rm H}_N(x;y) $ analogous to (\ref{rho_exp}). Thus we seek the 
asymptotic expansion of the determinant specified in (\ref{ibg_Hdet}), 
(\ref{Hmatrix}), which can be considered as a Hankel generalisation of a 
Fisher-Hartwig type Toeplitz determinant, or equivalently the asymptotic 
expansion of the random matrix average (\ref{rho_GUEint}). On this latter 
viewpoint of the problem, we recall that the original Szeg\"o theorem for the
asymptotic form of the Toeplitz determinants with smooth, positive symbols has
been proven by Johansson \cite{Jo_1988} starting from the analogue of the random
matrix average (\ref{rho_Uint}) and then generalised to an analogous
theorem for Hankel determinants relating to Jacobi averages
\cite{Jo_1997}. We note too that some mappings of Fisher-Hartwig
symbols in the Toeplitz case to analogous symbols in the Hankel case are
known \cite{BE_2002}. 

Another aspect of the present work which provides the beginning for future
studies is our derivation of the recurrence relation (\ref{refl_diff}) using
orthogonal polynomial theory. As we have commented, the recurrence obtained 
via this method coincides with recurrence obtained by Adler and van Moerbeke
using methods from soliton theory. Of course one would like to understand
the underlying reason for this coincidence. Also, recent works of Borodin
and Boyarchenko \cite{Bo_2001}, \cite{BB_2002}, starting from a formulation
in terms of the discrete Riemann-Hilbert problem, provides alternative 
recurrences involving discrete Painlev\'e equations for closely related 
Toeplitz determinants. It remains to understand the relationship between the
different approaches. 

{\it Note added}: Subsequent to the completion of this work an asymptotic form
analogous to (\ref{rho_exp}) has been derived in \cite{FFGW_2002b} for the harmonic
well case, and from this it is deduced that, as with periodic boundary conditions, 
the $ \lambda_k $ for fixed $ k $ are proportional to $ \sqrt{N} $.

\begin{acknowledgement}
This research has been supported by the Australian Research Council.
NSW thanks V.~Korepin and A.~Fetter for comments and suggestions on this work.
\end{acknowledgement}

\bibliographystyle{cmp}
\bibliography{moment,random_matrices,nonlinear}

\end{document}